\documentclass[showpacs,preprintnumbers,aps,final,nofootinbib]{revtex4}

\begin{document}

\title{Is a classical Euclidean TOE reasonable?}

\author{A. Arbona}
\altaffiliation{Blanes 21, Portals Nous 07180, Calvi\`a,
Mallorca, Spain.}
\email{toniarbona@hotmail.com}

\date{\today}

\begin{abstract}
We analyze both the feasibility and reasonableness of a classical Euclidean 
Theory of Everything (TOE), which we understand as a TOE based on an Euclidean
space and an absolute time over which deterministic models of particles and 
forces are built. The possible axiomatic complexity of a TOE in
such a framework is considered and compared to the complexity of the
assumptions underlying the Standard Model. Current approaches to
relevant (for our purposes) reformulations of Special Relativity, General 
Relativity, inertia models and Quantum Theory are summarized, and links between
some of these reformulations are exposed. A qualitative framework is suggested
for a research program on a classical Euclidean TOE. Within this framework
an underlying basis is suggested, in particular, for the Principle of
Relativity and Principle of Equivalence. A model for gravity as an inertial
phenomenon is proposed. Also, a basis for quantum indeterminacy and wave
function collapse is suggested in the framework.

\end{abstract}

\maketitle

\section{Introduction}

It is frequently discussed whether a Theory of Everything (TOE) is
feasible as an ultimate synthesis for the physical description of
nature. We will not address this issue here, but assume this approach
is indeed possible. Instead, here we want to speculate about the 
feasibility of a classical and Euclidean TOE, that we define as a TOE which 
is based on an Euclidean framework of space and an absolute time over which 
deterministic models of particles and forces are built.

Obviously, this approach goes against the accepted axiomatic base for Special
Relativity, General Relativity and Quantum Theory. Of course, the
experimental results related to these theories must be reproduced in such a 
possible TOE, but its axiomatic and interpretational base should be 
completely reevaluated if we
ever want to consider the classical Euclidean approach (called classical
approach frequently from now on). Basically, this 
would mean to go back to the interpretational standards of the XIXth
century and reject the conceptual revolutions of the XXth century
physics, including our modern idea of space-time, its geometry, and
the fundamental indeterministic character of the physical laws.

An important point to state before considering such a wild speculation is that
we should bind ourselves to Occam's razor\footnote{"Entia non sunt 
multiplicanda praeter necessitatem"}. 
If we ever discover that the classical approach is feasible but at 
the cost of an overwhelming complexity on the underlying axioms, we 
should not consider such an approach as reasonable. 
Only if the axiomatic basis is economical in
assumptions and parameters, in comparison with the Standard Model, we
would consider it as an alternative. Special Relativity, General
Relativity and Quantum Theory have been placed at the core of the
Standard Model precisely because of their axiomatic simplicity.
Any alternative model cannot pretend to substitute or reinterpret them 
without being similarly economical (or even more economical).

However, we should not overlook the fact that there is a profound
inconsistency at the core of the Standard Model: Quantum Theory
and General Relativity, as they are understood today, are incompatible
(even the relationship between Special Relativity and Quantum Theory
is not as smooth as desirable, as we will see).
The current attempts at solving this problem are indeed complex in 
their underlying assumptions. String Theory, for instance, which is
usually considered the main hope for solving the inconsistency, challenges 
Occam's razor with such a complexity that leads many researchers to
state that we only have a superficial grasp on its structure. In this
framework, the tolerable complexity of a classical approach would become
quite high.

All this stated, it would seem quite hopeless to achieve the goal of a
classical approach if it was not for significant advances in several models
and research programs which stand up to the Standard Model. We will
account for some of this models on sections~\ref{firstPillar} to
\ref{fourthPillar}.

This paper has two goals. The first is to be a short review of several
hypothesis and research programs dealing with the unsolved inconsistencies in
fundamental physics, as the incompatibility between Quantum Theory and
General Relativity, and with unsatisfactory (to a minority of
physicists, at least) axiomatic groundings for the well-accepted
fundamental principles of physics. 
The second goal is to try to provide a speculative framework where some of 
these alternative models could be coherently combined. This will help in 
offering a qualitative perspective on the feasibility of a classical 
Euclidean TOE.

\subsection{Outline}
The following sections are organised as follows: 
in section~\ref{holes} we review some of the most worrying mismatches
and open questions in the Standard Model. We have selected those that
better reflect, in our opinion, the weaknesses of the Standard Model
and the complexity which raises from patching it without reconsidering
a reinterpretation of its basis. In section~\ref{firstPillar} we 
shortly describe a non-standard approach to the underlying principles of
Special Relativity. This is to show that one can consider hypothesis
which imply a violation of Lorentz invariance, or of the universality of 
speed of light in vacuum, on special circumstances. 
In section~\ref{secondpillar} we will confront the standard
axiomatic grounding for General Relativity with an alternative grounding: 
gravity as an emergent phenomenon. This alternative dodges the problem of the
quantisation of gravity. Also in section~\ref{secondpillar} we will review 
the present status of the two non-standard alternative hypothesis for 
explaining inertia. 
In section~\ref{thirdPillar} we will discuss the Copenhagen approach
to Quantum Mechanics and the efforts to develop a deterministic and local
alternative, as well as other options. 
Also in section~\ref{thirdPillar} we will discuss the standard approach to Wave
Function Collapse and alternative hypothesis. In section~\ref{fourthPillar} we
discuss the problems originated by the supposed point-like nature of
particles. In section~\ref{alternative} we will explain why
we think it is relevant to put all these apparently disconnected
issues together as parts of a research program. We will propose a
speculation for a unifying explanation to the groundings of Quantum
Theory, gravity, inertia and Lorentz invariance. 
In section \ref{secConclusions} we will conclude by summarizing the 
assumptions one needs in order to seriously consider a classical approach 
and its relative complexity, in comparison with the Standard Model. In the
Appendices we will speculate further about a precise toy model for the
classical approach, about the emergence of spin in these kind of models
as well as about some qualitative experimental predictions of the approach.

\section{The holes in the Standard Model}
\label{holes}

In the following subsections we will introduce a selection of relevant
challenges that the Standard Model faces. They range from experimental
or observational challenges (dark energy, the weightlessness of the void) 
to theoretical (the quantum gravity deadlock, the origin of inertia, 
the conservation of energy) and interpretational (the quantum collapse).
Together, they pose a serious threat to the self-consistence of the
Standard Model and suggest that, rather than ad hoc patching, a complete
reevaluation of the basic postulates is reasonably in order.

\subsection{Unifying Quantum Field Theory and General Relativity}
\label{quantumGravity}

The efforts towards the unification of General Relativity and Quantum Field 
Theory (QFT) date back to the 40's\footnote{See, for instance, the review by
Thiemann~\cite{thi01} and references therein, to appear in the 
\textit{Living Reviews}~\cite{LVR}.}.
Despite a great deal of effort invested in this quest, the goal looks 
still far from reach. It is not strange that doubts on the feasibility 
and sensibility of this unification have risen. 

In fact, not everyone is convinced that we need to quantize General Relativity
in order to unify it with Quantum Theory. Feynman, for instance~\cite{FEY95}, 
suggested that maybe gravity does not need to be quantized since gravitation 
can be the source of quantum collapse itself. This possibility has been
further explored by Fivel~\cite{FIV97}. Fivel notices that assuming there is
a fundamental force driving collapse, this force is at least as weak
as gravity. 
 
In a different direction, since Sakharov~\cite{SAK68} there have been many
approaches to gravity as an \textit{emergent}\footnote{We use emergent
  in the sense that the phenomenon emerges from other more basic
  interactions, as the Van der Waals forces emerging from electromagnetism, 
  for instance. We will repeatedly use (maybe even abuse) this term in this 
  paper to refer to such phenomena that are not fundamental but derived from
  others. This is to be opposed to fundamental phenomena that are
  grounded on irreducible axioms or assumed postulates.} 
phenomenon. Recent results~\cite{BAR01} show that possibly
gravity-like interactions are the natural long-distance effects of
most reasonable quantum fields. Also in the spirit of deriving gravity
from Quantum Theory, it has been suggested that the quantum void is the
origin of gravity~\cite{HAI02}.

An alternative research program has been summoning the efforts of many
theoretical physicists in the last two decades: String Theory. One of
the main reasons to pursue String Theory disregarding its lack of
contact with actual experiments is that it provides a mathematical
frame where gravity could be quantized. There are many introductory good
reviews on String Theory~\cite{GIB98, SCH00} so we will not describe the
string approach on the problems we will consider. Furthermore, we can
consider that what we
are seeking here is to collect and review a collection of hypothesis,
toy models and research programs that together could make up for an
alternative to String Theory as the next step in fundamental
physics. 

It is generally accepted, anyway, that the solution to the fundamental
incompatibility between QFT and General Relativity will imply a profound 
reshaping of the principles of at least one of those two fundamental theories. 
In the proposed framework on section~\ref{alternative} the fundamental 
principles in both theories are questioned. The standard principles are 
recycled as emergent laws with just statistical or long-distance sense.

\subsection{A dynamics for Wave Function Collapse}
\label{wfcDynamics}

The former problem of incompatibility is not usually bound with
the phenomenon of Wave Function Collapse (WFC) in a basic manner, if
we exclude the Feynman-Fivel remark (see previous subsection)\footnote{One 
may also consider Pearle (see subsection~\ref{pearle}) as a proponent of a
relationship.}, but for reasons 
that will be later explained, we will include it in the list of problems. 
The WFC phenomenon as interpreted and postulated in Quantum Theory is not a 
problem from the scientific point of view, since it is rooted as one of the 
axioms of Quantum Theory, and remains consistent with all observations up to 
date. However, it has been considered as a philosophical or interpretational 
problem by many renowned physicists, starting with Einstein~\cite{FEY65, PEN86,
PEA89}. One of the circumstances that better show the strange
behaviour of matter under WFC is the EPR phenomenon~\cite{epr},
which will be one of the basic ged\"anken test beds for our 
qualitative framework.

The efforts in order to construct a theory that provides a dynamical
rather than axiomatic explanation for WFC have been frequent
\cite{GHI86, GHI90, GHI95, PEA89}. Considering that the WFC
axiom is tightly linked to the non-deterministic nature of Quantum
Theory, the hypothetical achievement of the goal of a dynamical theory
for WFC would hopefully let us advance in a program to get rid of
indeterminism in the axioms of Quantum Theory~\cite{PEA89}.

\subsection{The origin of inertia}
\label{inertiaOrigins}

Another fundamental question that has been central in the development
of the modern gravitational theory is the origin of inertia. Since
Mach, there is a philosophical school that relates inertia to an
interaction with the most external shells of the Universe 
\cite{RAI81,WOO99,CIU95}. As it is well known, Einstein was heavily inspired 
by this relation in his works towards General Relativity. Inertia, in this 
view, is a gravitational effect.

We will see in section~\ref{secondpillar} that this approach is not still
generally accepted, and that there is a second school\footnote
{Of course, there is also the default assumption so that inertia is
actually fundamental and intrinsic to bodies.} that
bounds inertia to the properties of the local void~\cite{DOB00,RUE01,
HAI01,HAI02}.
This concept seems radically opposed to the Mach-Einstein line of thought, 
although the differences shorten if we link the local void with a stochastic
background (a sort of ether) in thermal equilibrium with the external
shells of the Universe. This second point of view on inertia will play
a central role in our discussion. 

\subsection{The void: just a Zero Point Field?}

Regarding to this \textit{ether}, a question that is sometimes related to 
the problems of inertia and gravity is the origin of the void, in the sense 
of the source of such phenomena as the Casimir effect or spontaneous
emission. The void is modelled in Quantum Theory as a Zero-Point Field
(ZPF), and this approach is in agreement with all experiments up to
date. 

From this point of view, the ZPF derives from Quantum Theory. However, 
the opposite view is gaining adepts: to derive Quantum Theory from the
ZPF phenomenology~\cite{IBI96}. Along this way,
an alternative program, known as Stochastic Electrodynamics 
(SED)~\cite{BOY75,PEN96}, tries to reproduce all the electromagnetic void 
phenomenology without recurring to Quantum Theory. SED assumes the existence of
a random background wave noise filling the void. This approach addresses the 
problem of indeterminacy.

\subsection{The cosmological role of the void}

It is known, furthermore, that something is missing in our understanding 
of the influence of the quantum vacuum at cosmic ranges. At least three 
issues show that the Standard Model needs a patch. 

First, if the quantum vacuum is subject to the gravitational force, the
resultant energy would be many orders of magnitude (around 120!) higher than 
observed~\cite{RUG00}, which has been named the \textit{vacuum catastrophe}
\cite{ADL95}. It seems that not only gravity resists quantisation, but
quantum vacuum resists to the supposed universality of gravity. This
helps in raising the suspicion that maybe geometry is not the ultimate 
description of gravity.

Second, the expansion of the Universe has been found to include a
runaway term: the expansion is accelerating~\cite{PEB02}. This term is usually
attributed to a \textit{dark energy} of the void. Hence, the quantum vacuum
is subject to some kind of force that permits runaway solutions.
It is not clear how this fits in the Standard Model in general, and
in the model of the 4 forces of nature in particular. For this reason,
this apparent force is sometimes called \textit{quintessence}, the fifth
element. The possibility that we have missed a fundamental force in our 
description of nature is quite real.

Finally, the nature of the quantum vacuum seems to be changing along
with the expansion of the Universe. The fine structure constant has
been found to be diminishing when cosmological lapses of time are
considered~\cite{WEB01}. This challenges the view of the quantum vacuum as a
static given background on which the rest of physics operate. The
quantum vacuum is probably dynamic and its evolution is linked to 
the evolution of the observable Universe. This gives further credit to
the option of deriving Quantum Theory from the ZPF phenomenology.

\subsection{Special relativity and the conservation of energy}
The standard space-time of Special Relativity implies, through Noether's
theorem, that (in particular) the relativistic energy-momentum is conserved. 
However, the theoretical
and experimental support to this conservation law is far from encouraging.
To appreciate the weakness of the energy conservation law, we note that
it fails in some way in all scales: it fails at a cosmological range
(dark energy and inflation)~\cite{PEB02}; at microscopical ranges it 
continually fails, although not on average (the Heisenberg uncertainty); it 
even fails at its safest old stronghold, the macroscopic ranges, when 
acceleration-related phenomena are involved (radiation reaction)~\cite{FUL60}.

It is not then reasonable to think that conservation of energy is not
fundamental at all? However, as mentioned, we know through Noether's 
theorem that the
conservation does not come from the properties of particular theories of
fields and particles, but from the same structure of space-time. Thus the
question reverts to: is not then reasonable to think that we have seriously
misunderstood the fundamentals of space-time?\footnote{You may be tempted
to think that this point favours String Theory rather than the pursue of
a classical Euclidean space. Why this is not the case in our opinion is
detailed in section \ref{alternative}.}

Notice, also, than in two of the three scenarios (dark energy and radiation
reaction), acceleration and runaway solutions seem to loom. We will suggest
this is also true in the third scenario and this will drive us to a
reconsideration of which kind of space-time we need for a background and
what kind of theories we should build on this background. 

\subsection{The alternatives}

In the four following sections we will discuss the origin of
these limitations of the Standard Model in terms of four putative 
pillars for the modern description of nature: Special Relativity, 
General Relativity and inertia, Quantum Theory, and point-like 
particles.
We will also introduce alternatives that seem interesting \footnote{Not
all of these alternatives are consistent with each other. To build a
coherent comprehensive alternative, some choices must be done.}
in the spirit of the classical approach.

\section{The first pillar. The principles of Special Relativity}
\label{firstPillar}

As it is well known, Special Relativity is built on top of the
Principle of Relativity (Lorentz invariance, that is, the equations
driving physics on any inertial frame are invariant) plus invariance of 
$c$~\footnote{Which may be viewed as part of the invariance or not. We keep
it apart for reasons that will become evident.}. Nonetheless, this
approach, while being the way Einstein unleashed Special Relativity
and the way Special Relativity is always taught, is not the only
possibility. There is another point of view, which can be useful to
construct Special Relativity, as reviewed, for instance, by 
Bell~\cite{BEL76,BEL87}. We will call it Lorentz's approach. 
Lorentz's approach (see next subsection) is not as 
elegant and concise as Einstein's, but it proves useful in order to 
provide some insight to certain (academic) problems. What is more 
important for us is that
Lorentz's approach is not rooted on top of the Principle of Relativity
plus $c$ as a bound speed, which can then be considered as consequences of
other principles rather than the axiomatic roots of relativity. In this 
sense, Special Relativity could be considered as emergent.

Whether we prefer one of these approaches or the other is just a
matter of philosophical taste as long as the theory is considered
closed and finished (at least, as far as electromagnetism is
concerned). However, if Special Relativity was going to be affected by
some sort of new phenomena or theory, the distinction could prove
critical. Lorentz's approach would tolerate certain modifications of
Special Relativity affecting the Principle of Relativity or $c$
invariance, as far as we recover the predictions based on these principles 
in most situations. Einstein's approach would not, since the Principle of
Relativity is considered a given principle on top of which the rest of
physics is raised: a \textit{metaprinciple}~\cite{LIB02}.

In fact, the generalisation from electromagnetism to the rest of physics
by constituting relativity as a metaprinciple above all physics is
considered the great triumph of Special Relativity. It is? We should not rule 
out the possibility that the extension was too hasty: even when most physical
theories seem to bind to Lorentz invariance, there are parts of physics 
that could escape this principle (at least in the broad sense that includes
$c$ invariance), that would be then downgraded to an 
emerging phenomenon.

\subsection{The relativity of Larmor-Lorentz-Poincar\'e~\cite{BEL76}}

Take as a starting point the experimental fact that the speed of light
has a well-defined value in a certain reference system (given by the
far stars, for instance - the comoving frame of the Universe), together 
with the remaining electromagnetic
laws (summarized in the Maxwell equations). If one provides a simple
model of matter based on electromagnetism, then the basic results of
Special Relativity emerge naturally for this basic model. Maxwell
equations are valid in any other reference system and the speed of
light is the same on such reference systems. Rods physically contract and
matter slows down its processes. Lorentz transformations show up. 
Bell calls it the Larmor-Lorentz-Poincar\'e approach and we refer
it as Lorentz's approach for short\footnote{For a complementary and
very interesting summary of the historical process of the rise and
fall of the electromagnetic TOE, see chapter 28 on Feynman's lectures
~\cite{FEY64}.}. 
Of course, one has to carefully reproduce the analysis for more
complicated physical systems; for instance, those involving nuclear
forces or radiation reaction. 

On the other hand, the Principle of Relativity takes the experimental
facts as postulates, and then the proofs get simpler and much more
elegant. 

Nevertheless, any fact of physics forces us to adopt one view or
another yet. Einstein's approach is clearly more useful, has an elegant
mathematical generalisation  and then it has
become the standard axiomatic explanation. However, we should not
overlook that if we ever observe a violation of the standard
postulates of relativity, Lorentz's approach could
cope with it under some circumstances, but Einstein's could not. That
also means that we should not always reject a hypothesis or model
based on the fact that violates the postulates of relativity, if the
violation appears on limited circumstances. 

\subsection{Violations of Special Relativity?}
\label{violatingRelativity}

We have already seen in the introduction that the standard space-time of
Special Relativity faces trouble when energy conservation is
considered. This suggests a possible deeper underlying basis for the
Minkowskian space-time.

Lately, furthermore, there has been an increasing theoretical evidence
that Einstein's approach is possibly too restrictive. 
To show this, we will basically follow the discussion by
Liberati, Sonego and Visser (LSV)\cite{LIB02}.
Special relevance is given to the Scharnhorst effect~\cite{SCH98} and to 
special spacetimes: the Wormhole, the Alcubierre warp drive and the
Krasnikov tube~\cite{LOB02}. Recently, it has also been suggested that the 
value of the speed of light (the fine structure constant, in fact) in the
vacuum may be decreasing in time, due to the influence of some sort of 
cosmological phenomenon~\cite{WEB01}\footnote{See Magueijo et al.~\cite{MAG02}
for a reason to derive
the $c$ variation from a fine structure ($\alpha$) variation.}. Finally, we
should also consider that there is an increasing feeling that phenomena
related to quantum entanglement reflect a conflict between Special
Relativity and Quantum Theory.

These evidences can be integrated into Special Relativity in two
ways. One can reformulate Special Relativity so that the Principle of
Relativity is preserved in a restricted sense, and renounce to invariance of
$c$. Alternatively, one can go further and recover Lorentz's approach to
Special Relativity, centred on electromagnetism or on any other theory with
the capability of being used as TOE. We bet for a variant of this
second option because it is one of the keys in order to get a
positive answer to the entitling question of this paper, as we will
discuss later. 

We will explain the
two possibilities, but first let us briefly introduce some of the mentioned
theoretical evidences.

\subsubsection{The Scharnhorst effect}
The Scharnhorst effect refers to a superluminal (faster than c) photon 
propagation in the Casimir vacuum (in a cavity with perfectly reflecting 
boundaries) due to higher order QED corrections. Certain frequencies 
are forbidden in the cavity, so diminishing the energy density of 
the vacuum. Therefore, if the energy density of unbounded vacuum is arbitrarily
taken to be zero, then the Casimir vacuum has negative energy density. 
This lower energy density induces, among other effects, a higher value 
for the speed of light. 

This is, by the way, another example of the difficulties that Special 
Relativity has with the concept of energy in some limits. 

\subsubsection{Special spacetimes}
There are a remarkable set of spacetimes which permit superluminal travel:
wormholes, the Alcubierre warp drive and the Krasnikov tube. All of them
violate the weak energy condition~\cite{LOB02}. 

Although classical forms of matter obey this energy condition, it is
nonetheless violated by certain quantum fields (for instance, in the Casimir
vacuum, as we have just mentioned). 

\subsection{Solutions}

The solution to these challenges can be approached from two radically 
different positions:

\subsubsection{Approach 1: Maintain the Principle of Relativity}

Now, the principle-centred approach to these challenges is as
follows. If one assumes as postulates the 3 following hypothesis:

\begin{enumerate}
\item In an inertial frame, the space is homogeneous, isotropic, and
          Euclidean, and the time is homogeneous

\item The Principle of Relativity
\item Precausality (causality in the standard sense, but without any
reference to coordinate systems) 
\end{enumerate}

then the existence of the Lorentz group is implied, containing a
parameter representing an invariant speed~\cite{LIB02}.

This does not imply an upper bound for the speed. However, given a
propagation phenomenon (like a wave-like propagation on a fluid with a
certain sound speed, which plays the role of an invariant speed) in
one reference frame, one can always find time and space coordinates,
in any other frame, such that the Principle of Relativity is satisfied
and an analogous equation holds true in the new frame, as LSV state. 

Once a finite value for $c$ is found experimentally, the whole formalism
of Special Relativity follows on the basis of postulates 1-3.

LSV note that one can always choose the value of $c$, because such a
choice is essentially equivalent to giving the prescription for
synchronization. However, the natural choice (so that the laws become
simpler, that is, Lorentz invariant) is to take the \textit{sound speed} 
as the synchronization speed. It
is easy to see that if the coordinates in two reference frames are to be
related by a Lorentz transformation with invariant speed $c$,
synchronization must be performed using signals that travel 
at the
speed $c$. For synchronization, then, one must use signals travelling at the
invariant speed $c$, if one wants to satisfy the Principle of
Relativity.

In this formulation, Special Relativity can tolerate
superluminal signalling at the kinematical level. There is nothing in
postulates 1-3 that implies a maximum speed. The existence of a speed
bound has to be added as an experimental input, or equivalently, as
an additional property of the laws of nature. If we rely on the
calculations of QED, we have to admit that $c$ is not such a bound, but
a sort of sound speed that characterizes for some reason all the known
laws of nature\footnote{This is true while one does not 
consider WFC as a dynamical phenomenon.}, although this sound speed is 
not usually surpassed. 

There are some consistency issues that need to be addressed,
however. First, in the Scharnhorst effect light behaves in a
non-Lorentz invariant way. But this is only because the ground state
of the electromagnetic field is not Lorentz invariant (due to the
boundaries, the Casimir plates). 

Also, causality paradoxes might show up. LSV show this is not the case
in the Scharnhorst phenomenon. If one synchronizes with the invariant
speed, one can still send signals at a higher speed so that causal
effects may be reverted in order in some reference systems (tachyons
going back in time). No causal paradox arises, though, if one can find a
preferred reference frame which defines the absolute time-ordering of the 
causal influences. Imagine $a$ causes $b$ through a tachyonic
signal in a certain reference system. There are other reference
systems where $b$ happens before $a$. The deep reason for this is,
however, that we build those reference systems by using
synchronisation at a finite speed that can be surpassed. There is no
causal paradox in that if one admits there can be a preferred
reference system. In the Scharnhorst effect there is such preferred
system: the rest frame of the Casimir plates. Coordinates are
arbitrary. To choose a reference system where the effect happens
before the cause is just a bad decision for a coordinate system, but
does not pose a paradox by itself.

\subsubsection{Approach 2: Lorentz relativity}

\textit{A natural step would be to reconsider the role of Lorentz invariance,
regarding it as a symmetry property of specific theories rather than
as a fundamental meta-principle of all physics}~\cite{LIB02}. As long as those
specific theories are just approximations to a more accurate physical
theory, violations of Special Relativity can emerge. 

So one can even dismiss the postulate 2: the Principle of Relativity. This
is Bell's and Lorentz's point of view. Lorentz invariance is then a 
property of electromagnetism. And a more profound understanding of 
electromagnetism could imply that Lorentz invariant is not completely general.

This second approach has the disadvantage of needing an explanation
for Lorentz invariance as a quite generalised principle, embodied in several 
laws of physics. Historically, Lorentz realised
of this. His approach was to consider electromagnetism as the fundamental
theory of nature and to build models of matter (electrons in particular)
based on pure electromagnetism. The electromagnetic picture of the world 
\cite{BEL76} is indeed an interesting academic exercise, but does not solve 
the problem as this picture has been abandoned as a possible TOE\footnote{We 
recommend again Feynman's lectures~\cite{FEY64}.}.

We need then to provide a fundamental reason for the coincidence in several 
laws of nature at approximately fulfilling Lorentz invariance (If 
electromagnetism is not the TOE we might need to find another candidate 
characterised with a sound speed $c$). We will try to provide such 
a reason in subsection~\ref{lorentzInvarianceOrigin}. 

The advantages of the approach, on the other hand, are the a priori absence of
paradoxes provoked by the Scharnhorst phenomena and relatives or by certain 
spacetimes, as well as the possibility of finding a cure for the
inconsistencies of Quantum Theory and energy conservation, as we will later 
recall.

\section{The second pillar. The Principle of Equivalence, gravity 
and geometry.}
\label{secondpillar}

The Principle of Equivalence is one of the pillars of General Relativity. 
Even most of the theories that have been proposed as an alternative to 
General Relativity attach
to this principle. There are good reasons for that, but the main one is
that the refutation experiments~\cite{ROL64, BRA72, ADE94} have established
that the principle is fulfilled with astonishing fractional
exactitude: a part in $10^{13}$~\cite{NOR02}.

It is clear, then, that the same force or pseudo-force that accounts
for inertial forces is also at the rooting of gravitational forces. The 
relation is necessarily profound, although when we turn to numbers and try to
derive inertia from General Relativity (as an emergent phenomenon) the proof 
faces important challenges, as we shall see in subsection~\ref{inertia}. 
The reason could be that even when they are related, inertia cannot be 
fully inferred from gravity because they both emerge from a more fundamental
force, or even that gravity should be derived from inertia and not the other
way round. This drives us to the question of whether General Relativity is 
itself a fundamental or an emergent phenomenon. 

Einstein showed that gravity can be understood as geometry. This
heavily points in the direction of gravity as a fundamental
phenomenon. However, we should remark that certain forces that act
homogeneously over mass have this capability of being described
through geometry (we will analyse this in the next subsection). In
other words, again we found a causal dilemma in the founding
principles: is gravity on the first place just a geometrical
phenomenon, or gravity is a special force which happens to interact
homogeneously with inertial mass and can be described, for elegance
and simplicity, formally as a geometry? The distinction would be
pointless, matter of taste again, if General Relativity were a final theory. 
However, the incompatibility between General Relativity and QFT is generally 
attacked via the quantization of General Relativity. If gravity is strictly 
geometry, we need then to build a quantum geometry, which is something that 
has been intensely tried for decades without success (although many theorists 
believe that String Theory will permit it). On the other hand, if geometry is
just an elegant mathematical shortcut for most situations in
gravity-conducted events, there is no point in pursuing a quantum
geometry.

\subsection{Gravity as an emergent phenomenon}
\label{emergentGravity}
Regarding geometric-like forces, we can recall the fact that the
refraction coefficients in materials, and so the curved paths of light
in those materials, can be described from the phenomenological point
of view as a geometry~\cite{GOR23}. That works at large scale, although if we
dive into the microscopic machinery, the geometrical description
cannot explain all the details of the behaviour of light. 

This can be directly related to the efforts in analogue models of General 
Relativity
that have been recently attracting much interest. Since actual experiments on
intense gravitational fields are not possible today, some physicists
started developing analogue models of General Relativity in the frame of other
branches of physics: acoustics, optics, and solid state, essentially. The idea 
is to simulate some features of General Relativity (as black holes) within a 
system in a lab~\cite{VIS01}. 
They have shown that the internal forces in these systems, 
while acting on a certain wave function (as light) can be interpreted as a
metric. Furthermore, Barcel\'o, Visser and Liberati (BVL) have recently
shown that one-loop effects contain a dynamics over that metric, in
the same geometrodynamical spirit of General Relativity~\cite{BAR01}. This 
approach is partially similar to
Sakharov's~\cite{SAK68}. Sakharov found out that if we renormalize on a
curved space rather than in flat space there appear remaining terms
in the action. The first of them is formally identical to that of
General Relativity. The rest he called them quantum corrections to General 
Relativity. Sakharov
provides a hypothesis for the dynamics of curved space but not for the
original reason of curved space itself. The BVL approach is more
profound, in our opinion, since the metric is not arbitrarily provided
but derived from the (Euclidean) physics of the system. In this sense 
General Relativity 
can be dubbed as an emergent phenomenon, since emerges as a long-distance
effect of other more fundamental forces. If so, this approach could play
a role on a classical approach: it permits us to recover Euclidean geometry 
as the underlying geometry. 

A real difference between the analogue models and gravity is the
strong coupling between this force and inertia. The simple example
with refraction just works with light, but fails to explain the
behaviour of other kind of waves. It fails because the interaction
does not couple to inertial mass, but to other properties of the
waves. More sophisticated models also have to battle against different
metrics obtained for different fields (which can be called
birefringence), although we will later suggest that this can turn into
an advantage (see section~\ref{explainepr}). Gravity, on the other hand, 
does couple to inertial mass. This is stated by the Principle of 
Equivalence, and has been experimentally shown to be highly 
accurate, as it was previously remarked. 

\subsection{The roots of inertia}
\label{inertia}

\subsubsection{Inertia and Mach's principle}
Let us jump then to the question of inertia and its relation to
gravity. The problem of inertia was one of the main reasons that drove
Einstein to the theory of General Relativity. The issue was known 
long ago, and was popularised by Mach. To summarize it, the 
problem comes from the fact that inertia is a force that depends on 
local acceleration. However, how does one establish a reference to which 
a system does not accelerate, or equivalently, how can we define 
absolute acceleration?

First, a precision is in order regarding the Mach's principle. This
principle embodies a much wider collection of issues. It has 
been so used and reinterpreted that is presently blurred enough to provide 
conflicting predictions on certain scenarios. The account of Bondi and Samuel
\cite{BON96} provides a listing of eleven different formulations of
the Mach's principle. The first of them (Mach0), \textit{Local inertial frames
do not rotate relatively to the Universe}, is in fact an experimental
observation rather than a principle, and then is what cosmological
theories are confronted with. In this sense, \textit{Mach's principle
is the experimental observation that the inertial frame defined by
local physics (zero Sagnac shift) coincides with the frame in which
the distant objects are at rest}~\cite{BON96}. The Sagnac effect is the
modern version of the Newton's bucket. When other variations of the
principle are considered, then theories can be considered Machian or
Anti-Machian depending on its fulfillment of the particular version of
the Mach's principle.

Following Mach, Einstein developed a theory that would explain inertia
as an interaction with the far stars. With this, one can introduce a
reference system that does not accelerate with respect to the far
stars, so one can refer to absolute accelerations and then make
sense of inertia. However, when we try to calculate in detail how
inertia emerges from a gravitational interaction with the farthest
regions of the Universe, several problems arise. But let us first describe 
the three options where all accounts of inertia can be classified into.

\subsection{Theories of inertia}

The first option is that inertia is an intrinsic property of massive bodies. 
This is simple and self-consistent\footnote{Although quite peculiar when
combined with Special Relativity: Special Relativity does not permit an
absolute space in terms of speed-related effects. Intrinsic inertia, on the 
other hand, requires an absolute space in terms of acceleration. This point of
view may be valid but it rather makes us feel there is something we do not
properly understand about the structure of space.}, and the default
explanation if we cannot produce anything better~\cite{DOB00}. However, if 
we can explain inertia
by other means, we then remove an unnecessary and strange postulate, which 
simplifies the theory. Thus it is worth to take a look at the alternatives.

The second option is that gravity is the source of inertia, as Einstein
(initially) believed. In fact, it has been shown that General Relativity
implies that the external shells of the Universe interact with local
(from our solar system, for instance) matter with a force that is
proportional to acceleration~\cite{WOO99, RAI81}. Several problems remain, 
however. First, if we want to honour the Principle of Equivalence, the factor 
in that proportion needs to be somewhat artificially adjusted. Second, the
instantaneous character of the inertial force needs some explanation,
since the external shells of the Universe are quite away, but the
gravitational reaction force, which is supposed to account for inertia, is
doubtlessly instantaneous. 

This paradox is very serious and the possible solutions within the orthodox
view are all quite unnatural. One possibility, defended by Ciufolini and
Wheeler~\cite{CIU95}, for instance, is that the force is propagated instantly.
This is a very radical and complex reconsideration of the Standard Model. 
It relies on the fact
that the Einstein equations contain both hyperbolic evolution equations (6 of
them) and elliptic equations (4 of them), which are in fact constraints on the
evolution of the system. The equations can only describe the coherent 
evolution of the metric and the matter (the left-hand and right-hand sides 
of the Einstein equations) and then we cannot externally alter the evolution of
matter at will: the constraints must be respected. This in turn puts
constraints on the evolution equations of matter. So if this program can
be brought to and end, it tells us that inertia must be implicit in the
evolution equations of matter\footnote{Unless gravity is shown to explain the
equations of matter. This means building a gravitational TOE, which
Einstein unsuccessfully tried for 30 years.}: that is not, in fact, anything 
very different
from the point of view that inertia is intrinsic to matter, unless we take
indeed the constraints as really decoupled from matter and being carriers of a 
new non-local force. In the first case the postulate of intrinsic inertia
is maintained and it is hard to defend that we have saved any complexity.
The second option is a quite ad hoc postulate with very important side
effects on the rest of physics. Non-local forces introduce a deep level
of complexity in the description of nature.

Another possibility within the gravitational origin of inertia is that
defended by Woodward and Mahood. According to their theory, a local
acceleration creates a wave which travels to the ends of the Universe,
makes it wiggle so that it creates a return wave which travels back in
time towards the source of the perturbation and interacts with it causing
inertia. This mechanism is recycled from a theory developed by Dirac,
Feynman and Wheeler, which they hoped was a response to the inconsistencies
of point-like electrodynamics. The theory was abandoned long ago in
electrodynamics, but in General Relativity is defended as the last hope
for preserving Einstein's view that the theory would explain inertia.
Dobyns, Rueda and Haisch have a good account that details why this possibility
is not very satisfactory~\cite{DOB00}. We will give more details about it in
the conclusions of this subsection. Also, this orthodox approach goes 
against our proposed paradigm\footnote{Even without considering the particular
use of the macroscopic (or rather cosmological) travel to the past, whose
consequences and possible paradoxes if generalized to all physics can be hardly
seen as a simplification with respect to the default explanation of intrinsic
inertia.}, as we will show, so we adhere to the refusal.

The third option is an alternative hypothesis has been recently concentrating 
some efforts. The hypothesis is still not a finished theory but a framework in
construction. The main point in the hypothesis is that inertia is the
result of a local interaction of matter with the quantum Zero-Point
Field (ZPF)~\cite{HAI01, RUE01, HAI02}. The advantage of the hypothesis is 
that circumvents one of the two problems above: the interaction is local. The 
difficulty in explaining the factor in the force is maintained. This and the 
fine-tuning of factors are completely different from the gravitational 
theory, though. We attach to a variant of this point of view for reasons that 
we will comment in section \ref{alternative}.

\subsubsection{Summary: gravity = inertia + geometrodynamics
(general relativity)}

The Principle of Equivalence simply states that gravitational forces are
equivalent to inertial reaction forces. But that does not mean that either
the Principle of Equivalence or General Relativity explain inertia. To 
affirm that General Relativity is the origin of inertia is another
completely different hypothesis that tries to close the loop, and which
should be separately considered. In fact, the claim that inertia is a
geometrodynamical effect is quite circular~\cite{DOB00}.

The alternative and consistent view in order to explain the Principle of
Equivalence is that, once inertia is externally (through an interaction 
with the ZPF, for instance) or intrinsically given, if the spacetime is curved 
then gravitational forces are in fact inertial reaction forces. What General 
Relativity does is to assume the space is curved and to provide a dynamics 
for it (the geometrodynamics), but that does not explain neither inertia nor 
the gravitational forces (which are the same thing).

In fact, the attempts at explaining inertial forces in the framework of
General Relativity are attempts to derive forces on matter generated by
geometrodynamics, which is quite paradoxical. One of the approaches is to
use the linearisation of General Relativity, which is the choice of
Woodward of Mahood based on an argument by Sciama (used by Sciama to
refute General Relativity, by the way). Another approach is to consider
that the elliptic constraints of geometrodynamics generate forces on
matter, as we have discussed.

The confusion comes when we do not distinguish gravitational forces
(weight) from geometrodynamics. Weight is the resistance of a body to
be separated from its geodesic, which is nothing else than inertia. But
nothing in geometrodynamics can explain the force that bodies feel when
they separate from geodesics, because geometrodynamics just describes
geodesics.

In conclusion, inertia in flat spacetime is weight in curved spacetime.
General Relativity explains neither, but only how geodesics evolve. Why
test particles feel forces when they deviate from geodesics is something
that only a general theory of inertia can explain.

\subsection{The current views on inertia and gravity}

In the last two subsections we have addressed the current approaches to
gravity and inertia separately. Now we will integrate them into combined
theories of gravity and inertia.

We have explained that the phenomenon of gravity is the sum of 
geometrodynamics (General Relativity) plus inertia. A first family of
theories (Woodward and Mahood, Ciufolini and Wheeler) speculate that
General Relativity as dynamics of geometry is an almost final theory (only
its quantisation pending) and
additionally a theory of inertia. While the first statement can be
defendable, the second is quite problematic, or even paradoxical.

A second option, the conservative approach in the Standard Model, is that
General Relativity is a final theory, which describes the dynamics of
geometry, and that inertia is an intrinsic property of bodies. This
approach is perfectly consistent, if we disregard the problems of
General Relativity with QFT.

The third option (Sakharov, BVL) states that geometrodynamics is an
effective phenomenon created by one loop effects of QFT. The underlying
spacetime is Minkowskian. Inertia is also intrinsic to bodies in principle.

The fourth option (Haisch, Rueda, Puthoff) is that inertia emerges from
an interaction between the ZPF and test particles. There is a natural
tendency in these theories to also consider geometrodynamics as emerging
from the ZPF, although this second step is more involved than the first. 

\section{The third pillar. Quantum theory: indeterminism and non-local 
collapse}
\label{thirdPillar}

There are two aspects of Quantum Theory that many scientists find hard
to swallow: its axiomatic or fundamental indeterminism and the vague, if
consistent, measure process related to the collapse of the wave function,
with its associated implicit non-locality.

For these reasons, many physicists have considered the theory as 
a provisional phenomenological construction. Its axioms, certainly vague
and questionably self-consistent, should not be seriously taken without 
further analysis. The precision of the predictions, however, drove the
orthodox view, heralded by Bohr, to ignore such considerations in
favor of the apparently simplest approach: to take the axioms as given 
metaprinciples, and with them, rooting the fundamentals of physics on 
indeterminism and non-locality.

The scientific debate on both aspects is frequently clarified through
the discussion on the so-called EPR paradox.

\subsection{Indeterminism and non-locality: The EPR paradox}
The paradox of Einstein, Podolsky and Rosen (EPR)~\cite{epr} was proposed as an
argument to show that Quantum Mechanics could not be a complete theory,
but it should be supplied with additional variables. These additional
variables would then ideally restore the apparent lack of causality and 
locality in the theory. 

Rather than the original EPR argument, a more incisive version by Bohm and
Aharonov is normally used as the discussion arena~\cite{BHO57}.
The EPR argument \`a la Bohm-Aharonov is as follows: consider a pair of 
spin $1/2\hbar$ particles created
somehow in the state spin singlet and that move in opposite directions.
Stern-Gerlach magnets, for instance, are provided to obtain measures.
If when measuring the projection of one of the electrons we obtain $+1/2$,
as the spin projection, then the measure of the other device must return 
$-1/2$, and vice versa. This
follows unquestionably from Quantum Mechanics and does not depend on
the distance between the two measuring devices, that can be large
enough to discard non-superluminal influences.
As we can predict the outcome of the measure on any component of the
second spin, then the result of the second measure must be predetermined. As 
the original quantum wave state does not determine the result of the first
measure, this predeterminacy suggests the possibility of a more complete
specification of the original state. 

The standard interpretation assumes alternatively that the first of the 
two EPR observations fixes what it was until then unfixed, not only in 
the first particle but in the second, which raised well-founded concerns 
about the locality of QFT. To EPR this was a \textit{spooky action at a 
distance}. In fact, as Bell stresses~\cite{BEL87},
Einstein was more concerned on non-locality rather than on indeterminism.
What was sacred for him was the principle of local causality, or
\textit{no action at a distance}.
To avoid this spooky action, EPR and followers are forced to attribute, to the 
affected space-time, properties which were \textit{real} before the measure,
properties which are correlated and predetermine the result of these
observations. As these properties, fixed before the observation, are
not accounted in the quantum formalism, then to EPR the quantum formalism
is incomplete. 

\subsection{The orthodox interpretation: the subject-object duality}

Quantum Theory in the Copenhagen interpretation deals basically with 
observations. It divides the world in
two parts: an observed part and an observing part. Quantum Theory
always refers to an external observer (which is quite amazing, in particular,
when taking the cosmological perspective of the Universe as a whole). 

In fact, the orthodox response of Bohr to EPR
implies that we must consider the experimental device as a whole, and not 
try to divide it in pieces and analyse it, with indeterminacy quotas 
separately located as if the device was nothing but a complex quantum
system. The subject is \textit{classical}, the object is
\textit{quantic}, and they obey different laws.

Even when indeterminism could be questioned as part of the underlying
axioms, non-locality would be considerably trickier. The profound reason 
is that the theory has to do, basically, with the results of measures and then
it presupposes an externally given distinction between the system (or object) 
and the measurer (or subject). The interface between them is not supposed
to be dynamical, which permits non-local phenomena when the quantic
object is measured by a large enough classical subject.

\subsection{The problems of the orthodox view}

The orthodox view leaves many questions unanswered: are instantaneous the 
jumps? Is there any way to predict a collapse? What or who is a measurer? 
What is the frequency of collapses? On what depends? 
How do we describe the quantic-classical transition?

\subsubsection{The ill-defined triggering of the collapse process}

One of the most apparent non-localities of Quantum Theory is the instantaneous,
and global in space, Wave Function Collapse after the measure. Quantum
Theory does not provide any means to establish what triggers a collapse,
whether it is instantaneous, non-dynamical and non-local or, on the
contrary, a dynamical process. As the orthodox view solves these questions 
through the uncanny subject-object duality, the only answer is that it 
simply \textit{happens} when a \textit{measure} is performed.
However there is no simple way to define the concept of measure, which is 
vaguely placed as something that happens at the classical level, more or 
less driven by the observation (of a conscious being, to make things worse). 

The solution of the external conscious observer \`a la Copenhagen in order
to complete the theory is just so complex that should be abandoned on pure 
application of Occam's razor\footnote{Even when the axioms look simple.}. 
How can one define in a simple way, without 
considering the interaction of millions of atoms, the concept of 
consciousness? 
How can that be formallised in a solid mathematical way? 

So we are just left with suppositions about the nature of the collapse 
triggering.
\textit{
Shouldn't we necessarily admit that processes more or less of the sort of a 
measure happen more or less always, more or less everywhere? Is there
but an instant then in which the jumps do not exist and the Schr\"odinger
equation can be applied?}\cite{BEL87}. 

\subsubsection{Non-locality: the collapse and correlations}

The supposed non-dynamical collapse process, furthermore, introduces
the problem of non-locality and the confrontation with Special Relativity.
Taking EPR as a paradigm, we can see that
the logic drives us to a deadlock: the correlation in EPR implies
than the result on one of the experiments reveals the result of the
second experiment, whenever the analyzers are parallel. Quantum
correlations are not apparently explainable in a local way, which 
raises the need for the non-dynamical mysterious WFC. This would not only
be a long-distance eerie influence, but it would propagate faster
than light. 

In a theory where parameters are added to Quantum Mechanics in order
to determine the results of experimental measures, without changing the
statistical predictions, there must exist a mechanism so that the 
orientation of a measuring device can influence the result provided
by the other instrument, no matter how far it may be. Moreover, the 
involved signal must propagate instantly, and then the theory could
not be Lorentz invariant. Of course, if the predictions of Quantum
Mechanics have a limited validity, then the situation is different. In
particular a superluminal signal could be enough.

\subsection{Alternatives} 

\subsubsection{Decoherence}
Decoherence is the minimal modification to the Copenhagen approach in order
to remove some of the most obvious incongruences of this interpretation
\cite{JOO85}. 
By introducing decoherence, the conscious observer is replaced by a noisy
environment which triggers the measures and the collapses. The resultant
theory is still non-local and indeterministic.

\subsubsection{Theories of local hidden variables}
The relevant question in order to get rid of quantum indeterminism and
non-locality with minimal changes is whether a scheme of hidden 
variables with the desired local value (in the sense of respecting Special 
Relativity) can be found. 
Bell and followers showed that this is not possible, though, if one
is to reproduce all the results of Quantum Theory~\cite{BEL64,CLA69,WIG70}.
Some hoped that the predictions of Quantum Theory in conflict with the
local hidden variables theories would prove mistaken. However, the 
experiments showed this hope was vain. Quantum Theory is right and
the local hidden variable theories are wrong~\cite{ASP82}\footnote{There
are still many papers produced nowadays on possible flaws of Bell's 
conclusions,
although most researchers would agree that we should attach to those 
conclusions by now. We do not consider the possibility of building a classical 
approach that is partially based on a flaw of Bell's theorem, and that
obviously makes the path towards a classical approach more challenging.}.

Hidden local variables are then not compatible with the statistical 
predictions of Quantum Mechanics. It is the requirement of locality, or more 
precisely that the result of a measure on a system is not affected by 
operations on a distant system with which it has interacted in the past, 
what creates the strongest difficulty. 

\subsubsection{Non-local hidden variables}

Nonetheless, several hidden variables interpretations of the elementary 
Quantum Theory have been explicitly built since the seminal works of de
Broglie and Bohm~\cite{BHO52}. They match all the predictions of 
Quantum Theory. Of course, these interpretations have a profound non-local 
structure. The descriptions of de Broglie-Bohm and followers embody a quite 
remarkable characteristic: the consequences of the events in one place 
propagate faster than light to other places. At least in principle, therefore,
a cure can be found for indeterminism, but not for non-locality (in the sense 
of Special Relativity, we should stress).

The conflict raises then again when we have no choice but to admit that the 
equations for 
hidden variables have generally a non-local character. In many alternative 
theories, there exists a causal explicit mechanism so that the disposition 
of one of the measuring devices influences the results obtained on a 
distant device. In fact, the EPR paradox finds its solution in the 
form Einstein would have liked the less. 

As Bell tells us, one of the options to skip this non-dynamical WFC is, 
in fact, that causal influences propagate faster than light. The role 
of the Lorentz invariance in the complete theory would be then quite 
problematic (which is anyway not a burden from our point of view: see 
section~\ref{firstPillar}). 

This is then the second true problem of the Quantum Theory:
the manifestly essential conflict between any precise formulation and
Special Relativity. That is, we find ourselves facing an apparent
incompatibility, at the deepest level, between the two fundamental pillars of
the contemporary physical theory (the first and third ones, in our particular
listing). This piles on the incompatibility between the second and third
pillars.

\subsubsection{Dynamical collapse models}
\label{pearle}
Another development line towards a greater physical predictability would be 
to have a jump in the equations as a dynamical process in definite
conditions. The jumps violate the linearity of the Schr\"odinger
equation, so the new equation should be non-linear. 
In the same line as the orthodox proponents, the evolution of the statevector
should be modified but to do it so the collapse is not instantaneous but, 
rather, follows a well-defined dynamics of a modified Schr\"odinger equation.

Even without an explicit dynamical process, the theory can be made more
consistent.
A specific example of spontaneous WFC is specially simple
and elegant~\cite{GHI86}. The wave evolves according to the Schr\"odinger 
equation but it collapses from time to time. 
The collapses are fortuitous and spontaneous, mathematically represented
by a probability of collapse. This theory does not add variables. By 
adding mathematical precision to the jumps of the wave function, it makes
more precise the action at a distance of ordinary Quantum Mechanics, and
thus we get rid at least from the need of an observer. The
most molest aspect of the theory is again the difficulty in reconciliating it
with Lorentz invariance, which is not anyway an essential trouble for us, as
already mentioned (What we cannot accept if we attach to the classical
approach is strict non-locality, but downgrading Lorentz invariance poses
no problem). 

Even when this sort of models make the Quantum Theory more precise and less
self-inconsistent, they add a new postulated element, so increasing the
complexity of the Standard Model. An even better solution would be then to 
provide a reason for the probability of collapse.

\subsubsection{Dynamical collapse models with a chooser}

In order to describe more accurately the apparently random choices made by 
nature the Schr\"odinger equation must have a chooser in it. 
P. Pearle has a good review~\cite{PEA99} on formalisms and models which modify 
the Schr\"odinger's equation so that it describes a WFC as a dynamical 
physical process. One of the strongest motivations for this addition, as 
mentioned, is to avoid that either human beings or unknown choosers are 
essential to the description of nature. 

Pearle settled on modelling that
chooser by external random noise~\cite{PEA79}. With external random noise 
as the chooser, \textit{it turns out that the mechanism for obtaining 
agreement with the predictions of Quantum Theory is very simple, which 
suggests one is on the right path }\cite{PEA99}.

Collapse models should provide an explanation for the random results
which ocurr in nature and are unexplained by standard Quantum Theory: 
\textit{the results was this rather than that because the noise fluctuated this
way rather than that way}~\cite{PEA99}. What is left unexplained is what is 
this noise and how its fluctuations can be computed. Pearle thinks that small
perturbations in the gravitational field are the ultimate source of noise.
We commented on section \ref{holes} that Feynman and Fivel also defended
that point of view.

ZPF (in the SED sense of HRP) is a better candidate of noise chooser for 
us\footnote{We will also trace gravity as a consequence of the ZPF, so this 
is not necessarily an opposed view to the Feynman-Fivel-Pearle vision.}.
This is not paradoxical since we segregate the ZPF phenomenology from the
rest of QFT. In fact, we have already suggested that ZPF can be seen
as the root of QFT rather than the other way round. By building QFT on
top of the stochastic ZPF we are automatically provided with a source
of noise, and with it with a reason for the apparent indeterminism of
Quantum Theory. 

The fact that we cannot usually predict the outcome of a quantum system 
remains, as the complete knowledge on the details of the noise is impossible.
The noise plays the role of the hidden variables.
Analogously, the description of the Brownian motion, for instance, could
have been in principle developed in a purely statistical way, while
making the statistics later understandable with the hypothesis of the 
molecular constitution of fluids. The additional fact that the measure
perturbs the system (see next section) increases the difficulty to model
the underlying noise.

\subsubsection{Non-Lorentzian deterministic dynamical collapse models
with a chooser}

Although this later approach is valid and goes in the right direction for
the purposes of the classical approach, there is the possibility of
going further: to model a collapse which is not only well-defined 
(deterministic), but also dynamical (that is local). This is only possible, 
of course, at the price of sacrificing Special Relativity. We will try to 
provide a prescription for such a framework in section~\ref{alternative}.

\section{The fourth pillar. The problematic point-like particle}
\label{fourthPillar}

One of the burdens that sunk the interpretational paradigm of XIXth century 
physics was the failure of the point-like models for particles. QFT overcame 
this problem with the wave-particle duality and renormalization, 
which in opinion of many are gross hacks (from the interpretational point of
view). Renormalization has been one of the 
great steps forward in physics and has given us Quantum Electrodynamics (QED),
the most precise physical theory ever. Nonetheless, it lacks mathematical 
elegance and seems too complicated. The machinery of QFT permits to deal with 
point-like particles, but at the price of an ugly squeaking. The so-called
\textit{finite QED}~\cite{SCH95}, for instance, looks for a cure to this 
problem. However, up to now we do not really have a satisfactory way to
combine point-like charges and electrodynamics. 

In String Theory particles are not point-like, but they have a
structure, which avoids many problems and could allow the description
of many quantized forces, including gravity. As we advanced in the
introduction, however, we are not interested in this way, so we will
not give further details on this. 

Any time we try to ground the particle concept, problems arise. So a
valid question is: what do we need particles for? In fact, do they
exist at all?\footnote{Of course, existence is not a very well-defined 
physical term. What we mean is, it is necessary for physical models of nature
to be based on particles? Could they just be an emergent phenomenon and useful 
to describe certain situations in nature, but actually not fundamental?}.
If we look into QFT, it appears that the only 
deep reason for keeping the couple wave-particle is the WFC. Can one go
further than String Theory and avoid not only point-like structures
but also permanent structures? Particles are usually assumed a
fundamental phenomenon, but perhaps this is not the only way to build
a satisfactory model of nature. One could imagine an underlying
interaction that produces particle-like events (localized in time and
space) as an emergent phenomenon, which we can model without reference
to point-like structures. The WFC, in fact, can explain by itself
particle-like behaviours of the quantum waves.

This point of view was initially considered when Quantum Theory was
emerging. As Bell comments~\cite{BEL87}, Jordan, for instance, hold the point 
of view that particles do not truly exist. He tells us that Jordan emphatically 
declared that observations not only perturb what it is to be measured; 
they \textit{produce} it. Also Schr\"odinger tried to substitute the 
particles with waves~\cite{SCH}. He envisioned a theory
of Quantum Mechanics rid of particles, that would free the theory
from the collapse problem. We attach to the point of view of Jordan and 
Schr\"odinger as another key of the classical approach.




\section{An alternative framework}
\label{alternative}

In this section we will try to integrate some of the alternative hypothesis
described in the previous sections into a coherent alternative framework
for the classical interpretation of Special and General Relativity, and
Quantum Theory. Several assumption we make are, we believe, quite general
within a classical Euclidean TOE. Others are more questionable, but
nonetheless interesting in order to proceed further. We group this later 
assumptions under the name \textit{Spin Force Model} (SFM).

\subsection{Assumptions}
\label{assumptions}
Let us assume our observable Universe can be represented by a fluid\footnote{
In the broad sense of a continuous field locally described by a density and
kinematical parameters.} on an 
Euclidean space, governed by deterministic laws (in the form of partial 
differential equations) free of singularity-causing shocks. The space is
isotropic and homogeneous, and it is embodied with an absolute time 
($t$). 
We also assume that the blob of fluid that amounts for our observable Universe 
suffered an explosion and it is expanding against a (relative) void.
The expanding blob is locally in an approximate thermodynamic equilibrium 
regarding to internal (turbulence-driven) noise if we
consider small lapses of time (compared to the time since the explosion),
and if we ignore areas of high concentration of the fluid (which we will
associate to matter and radiation). This fluid in thermodynamic equilibrium
is assumed to be the vacuum of QFT (the ZPF). 
Let us call it \textit{ether} to distinguish it from the thinner void 
outside the blob.

We assume a conservation law for the fluid, so that we can describe it through
a density function $\rho(x^\mu)$. 
We further assume that the fundamental Lagrangian density ruling
the fluid dynamics only comprises $\rho(x^\mu)$, its first derivatives
$\partial_\mu \rho(x^\nu)$ and the relative acceleration that 
characterises the motion of the fluid, 
$\partial_\mu \mathbf{a}(x^\nu)$ \footnote{We should
not completely discard that higher derivatives are needed.}:

\begin{equation}
	\mathcal{L} (x^\mu) = \mathcal{L} (\rho(x^\mu),\partial_\nu \rho(x^\mu),
	\partial_\nu \mathbf{a}(x^\mu))
	\label{subthermalLagrangian}
\end{equation}

where we have introduced the vector of coordinates $x^\mu=(t,x^i),\ i=1..3$ 
which represents the absolute time plus 3 space coordinates\footnote{We 
follow the standard 
convention of distinguishing space coordinates with Latin indexes and space-time
coordinates with Greek indices. Notice, however, that here the aggregation of 
space and time has only a notational purpose. The space-time is not to be 
understood as Minkowskian.}.

At times, it will be
convenient to simplify the description of the dynamics of the system
through the introduction of an effective force, which we will call 
\textit{ether force}. So we will refer to the dynamics resulting from
the evolution of $\rho(x^i)$ and $\mathbf{a}(x^i)$ in the equation
(\ref{subthermalLagrangian})
as ether forces, for short, whenever those dynamics can be assimilated
to an effective force. This approach will provide us with the basic tools
to explain inertia and Special Relativity.

Within the SFM, by definition,
we still further assume that the underlying theory presents vortex-like
solutions that can supply the phenomenology we need to describe particles,
including possibly spin. Additionally, in the SFM the ether force is supposed 
to be spin-dependent\footnote{Then in the SFM the spin may need to be 
axiomatically introduced. However, we do not discard that this phenomenon 
can itself be derived from the model. See Appendix \ref{spinOrigin} for a 
speculation
on that possibility.}, that is, each spin component of a point in the fluid
only interacts with the same spin components of the neighbouring points.
This particular assumptions will let us obtain a seed for a theory
explaining Quantum Theory and the Principle of Equivalence.

If we attach to this point of view, then the four pillars need 
reinterpretation. 
Lorentz invariance should be explained as a property emerging from the 
fundamental form of (\ref{subthermalLagrangian}). Inertia should 
be the reflection of the stochastic average ether forces between ether and 
matter when there are relative accelerations. 
Gravity should amount for a long-distance effect of this same ether force. 
Quantum uncertainty would be explained as the effect of the random noise of 
the ether, which exists at a certain range in space. The WFC would be directly
inferred 
from the ether force, in a mechanism very close to inertia. Particles would 
not exist as permanent point-like structures, but just as high-density waves 
on the low-density ether. The particle properties would emerge from the 
collapse of these waves. 

At a cosmological level, a suitable present condition for the observable 
Universe is an expanding sphere with localized high-density regions of the 
fluid representing matter (stars, planets and dust, mainly). At a fixed time, 
these high-density regions are in equilibrium, at macroscopic ranges (defined 
below), with a homogeneous low-density fluid that fills all the space, which 
we have called ether for short,
although it has to be understood in the previous sense. As the sphere
expands, the density of the ether diminishes, so modifying the sound speed $c$ 
(increasing it, like in the Scharnhorst effect). 

\subsection{Dynamics on different ranges}
\label{ranges}

Let us point out, before we continue, that this model implies four
different ranges in space, with different physics on each. 

\subsubsection{Subthermal range}

The shorter range is that where the density, speed and acceleration of the 
fluid vary smoothly. Such a range, let us call it \textit{subthermal range}, 
must exist because we assumed well-behaved fields in our definition of the 
fluid model. The initial conditions must be smooth and the equations cannot 
bear shocks. The dynamics on this range are ruled by the fundamental Lagrangian
(\ref{subthermalLagrangian}).

The fundamental equations can be seen as providing the dynamics for 
$\mathbf{a}(x^i)$. The Cauchy problem would be completed by adding
the continuity equation and providing initial data for $\rho(x^i)$, 
$\mathbf{v}(x^i)$ (the speed), $\mathbf{a}(x^i)$ and $\mathbf{e}(x^i)$ 
(the $jerk$), where we define

\begin{equation}
 \mathbf{e}(x^i) \equiv \partial_t \mathbf{a}(x^i)
\end{equation}

\subsubsection{Thermal range}

A medium range is that where density, speeds and accelerations vary 
dramatically at neighbouring points, so that the fluid behaves as a 
stochastic background in thermal equilibrium. Let
us call it \textit{thermal range} (it would presumably correspond to the
Planck scale). The thermal range shares many properties with the
Brownian motion and with turbulent phases in fluid mechanics. Let's
define $T_{th}$ as the typical time scale for the fluctuations, and
$D_{th}$ as the typical spatial scale.

To describe the dynamics on this range, the fundamental subthermal Lagrangian
(\ref{subthermalLagrangian}) is not appropriate unless numerical simulation
is involved. For a mathematical treatment it is better to move to derived
quantities on a pseudo-thermodynamic level in order to deal with the
stochastic background, and then consider the dynamics of the perturbations.

We will consider the mean values of density ($<\rho>_{th}$), speed 
($<v>_{th}$), acceleration ($<a>_{th}$) and jerk ($<e>_{th}$) on the thermal
range. We assume the equilibrium of these values at the thermal range in a
certain neighbourhood.
That is, they are independent from $x^\mu$ in this neighbourhood. In the 
comoving reference system we have then that

\begin{eqnarray}
  <v>_{th}& =& 0,\\
  <a>_{th}& =& 0,\\
  <e>_{th}& =& 0
\end{eqnarray}

The description of the thermal background at this range completes with the 
specification of the remaining momentum orders for $\rho$, $v$, $a$ and $e$:

\begin{eqnarray}
  <\rho^2>_{th},...,<\rho^n>_{th}\\
  <v^2>_{th},...,<v^n>_{th}\\
  <a^2>_{th},...,<a^n>_{th}\\
  <e^2>_{th},...,<e^n>_{th}
\end{eqnarray}

which are also invariant under spatial and temporal variations in the 
macroscopic range (see below). The maximum order $n$ depends on the form
of (\ref{subthermalLagrangian}). These momenta, however, are not generally 
invariant under variations at cosmological ranges.

We can assume that the system is described through an
effective Lagrangian which describes the dynamics of the perturbations
on the static background described by the thermodynamic quantities.
That is, the dynamics of the system could be described by the mean
values of density, speed, acceleration, jerk, their standard deviations
and possibly higher order momentums of these three magnitudes\footnote
{No simplification emerges yet in considering a purely Gaussian
noise.}, as well as by the variables describing the perturbation. 
By defining the n-th order momentum of $X$ by $<X^n>$ we can formally
write

\begin{eqnarray}
	\label{thermalLagrangian2}
	\mathcal{L}_{th} = \mathcal{L}_{th}&(&\delta_{th} \rho(x^\mu),
	\partial_\nu \delta_{th} \rho(x^\mu), \partial_\nu 
	\delta_{th}\mathbf{a}(x^\mu),\\ \nonumber
	&&<\rho^n>_{th},<v^n>_{th},<a^n>_{th}, <e^n>_{th}, N(x^\mu) ),\ n=1..m
\end{eqnarray}

where $N(x^\mu)$ represents a source of noise, $\delta_{th}$ represents the 
deviations relative to the background means and $m$ is to be derived from the 
precise formulations of (\ref{subthermalLagrangian}). As we assume 
isotropy, these means are described by scalar variables.
 
Notice that to solve the Cauchy problem, (\ref{thermalLagrangian2}) needs
to be supplemented with initial conditions for the perturbations on the 
density ($\delta_{th}\rho(x^i)$), speed ($\delta_{th}\mathbf{v}(x^i)$), 
acceleration ($\delta_{th}\mathbf{a}(x^i)$) and jerk 
($\delta_{th}\mathbf{e}(x^i)$) of the fluid.

The stochastic thermodynamic background provides in this range
a reservoir of noise which could be described through thermodynamic 
quantities, in particular through a energy, temperature and entropy\footnote{
Notice, however, that energy in the standard sense, and with it temperature, 
entropy and other thermodynamic quantities, is not yet defined at this 
level. 
Here we have instead a pseudo-thermodynamics based on the conserved quantities 
implicit in~(\ref{subthermalLagrangian}). See subsection~\ref{conservationLaws}
for details.}. This noise will be interpreted as the source of the 
uncertainty in Quantum Theory. In particular it will provide the virtual
particles that characterize Quantum Field Theory. Then the equations for
$\delta_{th} X(x^\mu)$ are not deterministic when this noise is considered.

\subsubsection{Macroscopic range}

The broader range is that where the neighbourhoods where stable vortex
solutions are absent (the ether) can be seen as homogeneous (the 
constant density, speed and acceleration can be modelled as a simple
non-dynamical background, without the fluctuations that characterize the 
thermal range, from which an entirely new space-time will emerge) ignoring the 
details of the underlying noise, which can be then described through more 
appropriate derived quantities.
 
The assumed equilibrium requires that, across the ether, the integral standard 
deviation (and higher order momentums) of $\mathbf{a}(x^i)$ on typical volumes 
are small-enough so that they can be locally taken as null.  
We can call it \textit{macroscopic range}. 

The mean values of the kinematical quantities ($<X_{(e)}>_m$) are in the ether
the same as in the thermal range:

\begin{eqnarray}
  <v_{(e)}>_{m}& =& <v>_{th},\\
  <a_{(e)}>_{m}& =& <a>_{th},\\
  <e_{(e)}>_{m}& =& <e>_{th}
\end{eqnarray}

In the comoving frame:

\begin{eqnarray}
  <v_{(e)}>_{m}& =& 0,\\
  <a_{(e)}>_{m}& =& 0,\\
  <e_{(e)}>_{m}& =& 0
\end{eqnarray}

The higher momentum orders are also null:

\begin{eqnarray}
  <v_{(e)}^n>_{m}& =& 0,\\
  <a_{(e)}^n>_{m}& =& 0,\\
  <e_{(e)}^n>_{m}& =& 0
\end{eqnarray}

for any $n>1$.

On the other hand, the neighbourhoods of vortex solutions are assimilated 
to points described by certain parameters (the numbers of the particle)
and surrounded by standard unperturbed ether (although we will consider
later the case where the perturbations that particles cause on the surrounding
ether cannot be overlooked).
In the neighbourhood of a particle the macroscopic means have 
well-defined
values (in the comoving reference system), which differ in general from those 
of the ether:

\begin{eqnarray}
  <\rho_{(p)}^n>_{m}& \ne& <\rho_{(e)}^n>_{m},\\
  <v_{(p)}^n>_{m}& \ne& <v_{(e)}^n>_{m},\\
  <a_{(p)}^n>_{m}& \ne& <a_{(e)}^n>_{m},\\
  <e_{(p)}^n>_{m}& \ne& <e_{(e)}^n>_{m}
\end{eqnarray}

The detailed description of the particle involves a statistical description
of the thermal averages of the shape of the particle all over the region where
the particle is distributed. These functions are neither
isotropic nor homogeneous in general:

\begin{eqnarray}
  <\rho_{(p)}^n>_{th}(x^i)& \ne& <\rho_{(e)}^n>_{m},\\
  <v_{(p)}^n>_{th}(x^i)& \ne& <v_{(e)}^n>_{m},\\
  <a_{(p)}^n>_{th}(x^i)& \ne& <a_{(e)}^n>_{m},\\
  <e_{(p)}^n>_{th}(x^i)& \ne& <e_{(e)}^n>_{m}
\end{eqnarray}

for $x^i \in \mathcal X$, where $\mathcal X$ is the neighbourhood that 
comprises the particle and $x^i$ is expressed in the comoving reference
system, so that the functions do not depend on $t$.

And then the thermal Lagrangian for a certain particle becomes in a first
instance

\begin{eqnarray}
	\mathcal{L}_{th} = \mathcal{L}_{th}&(&\delta_{m}\rho_{(p)}(x^\mu),
	\delta_{m}\partial_\nu \rho_{(p)}(x^\mu),
	\delta_{m}\partial_\nu \mathbf{a_{(p)}}(x^\mu),\\ \nonumber
	&&<\rho_{(p)}^n>_{th}(x^i),<v_{(p)}^n>_{th}(x^i),
	<a_{(p)}^n>_{th}(x^i), <e_{(p)}^n>_{th}(x^i),\\ \nonumber
	&&<\rho_{(e)}^n>_{th}, <v_{(e)}^n>_{th},
	<a_{(e)}^n>_{th}, <e_{(e)}^n>_{th} ),
	\label{particleLagrangian}
\end{eqnarray}

for $n=1...m$ and $x^i \in \mathcal X$.

The dynamics in the macroscopic range for an isolated particle 
($\mathcal{L}_{m}$) comes from the interaction between 
its deviations relative to the background and the background. The mean 
effect is a minimisation of the relative acceleration between the 
particle and the ether, that is inertia. This can be described through
the relativistic kinematics. We will further explore the relationship between
the particle and the ether when we analyse the second pillar. There we will
consider which quantities can $\mathcal{L}_{m}$ depend on. The true dynamics 
in this approximation will come from the perturbation that particles exert on 
each other, which modify the equilibrium $<X_{(p)}>_{m}$ values.

As the isolated particles, the ether neighbourhoods are not dynamical.
They constitute the Minkowskian space-time.


\subsubsection{Concluding remarks on ranges}

We should still consider a fourth range: the cosmological one, but we will 
go back to this later. 

One can make an analogy between the former 3 ranges and the
equivalent electromagnetic ones in Brownian motion: the subthermal
would be that describing the interior of the molecules, where quantities
vary smoothly (except around the nucleus, perhaps). The thermal would
be that describing the atoms while in Brownian motion. Finally, the
macroscopic range would be that describing the fluid at such a scale
that all the matter fields become continuous. One can describe phonons there,
for instance, without paying attention to the real structure of the fluid
background. In the same way, three time ranges must exist. 

We will suggest that Special 
and General Relativity, inertia, and the standard energy-momentum conservation 
laws are approximate effective theories that work well on the macroscopic 
range but not in the finer ones. 
QFT will be suggested as a correct approximation to the
thermal range, and hence more fundamental than Special and General Relativity.
The clashes between these theories and QFT (see sections~\ref{firstPillar}
and~\ref{secondpillar}) should then be solved favourably
to QFT. That does not mean that we accept the axioms of QFT as truly 
fundamental, but just effective at the thermal range and wrong at the 
subthermal range. At the subthermal range we assume a classical Euclidean
description.

Let us see in more detail how a framework for all this can be raised.
First we will revisit the four pillars and provide an alternative 
interpretation for their groundings.

\subsection{The fourth pillar. Vortex dynamics}

We have already adopted the point of view that particles, in the sense of 
permanent and point-like structures, do not exist. If we restrict ourselves to 
the alternative framework we are describing, then we should provide mechanisms
for the emergence of fermions, bosons, and their interactions purely on the 
basics of fluid dynamics.
This is beyond the scope of this paper. Anyway, some speculations are provided
in the Appendix~\ref{spinOrigin} and some basic necessary features will be 
discussed below.

If we can assume that particles do not exist, there is no reason for waves to 
get concentrated in points. They concentrate (in a small region but not in a 
point) when a WFC happens (a metastable state) or in stable configurations 
such as atoms, but the rest of the time they should travel quite freely 
\textit{out of matter} (in the same sense as the free wave function of QFT) 
and stochastically fill the space. As we have assumed a singularity-free 
model, we can attach to this view.

Fermions should be then interpreted as vortices on the fluid. Generally,
vortex-like structures on a fluid with a well-defined sound speed will
propagate slower than phonons (which propagate typically at $c$, the sound
speed), as they keep part of 
their structure while traveling (which invests part of the speed in internal 
rotation, that is, keeping its spin\footnote{In the broad sense of a permanent
or characteristic internal rotation that represents a footprint of the vortex. 
It may happen that the quantic spin do emerges from this rotation, as we 
will comment in the Appendix \ref{spinOrigin}.} properties).

In general, fundamental bosons could be mapped to macroscopic-range phonons 
travelling through the background fluid at the sound speed, while fermions and 
derived bosons (those composed of fermions) would keep a permanent 
internal rotation, so travelling at subsonic speeds. Supersonic speeds are
also possible when the macroscopic range approximation is not valid.

Inertial mass would be derived from the interaction of the internal
rotation of these vortices, which necessarily create an acceleration
field, with the corresponding spin components of the
surrounding ether. The specific value of inertial mass should be computed
from the dynamics of the fluid vortex lines, based on the fundamental fact
that ether forces emerge when there are relative accelerations. It is 
presumably very difficult to compute analytically these values from most 
models deriving from (\ref{subthermalLagrangian}). Numerical simulation should
prove a more practical tool.

The long-range effect of these vortex structures must necessarily depend on
this same effect between internal rotation (spin) and surrounding ether,
as we must recover the Principle of Equivalence. We will come back to this
when discussing the second pillar.

\subsection{The first pillar. Acoustic relativity}

As advanced in the introduction of this section, we understand Special 
Relativity as an effective theory valid on the macroscopic range. On this 
range, the ether is seen as homogeneous and
isotropic. Causality is always granted as the underlying theory provides
an absolute time. No paradoxes related to causality violation can appear
neither they need to be separately considered. To obtain Special Relativity, 
as we have explained in section~\ref{secondpillar}, we only need to provide 
a reason for Lorentz invariance and a sound speed.

\subsubsection{Speed of light}

The sound speed is given by the propagation speed of macroscopic-range
waves (phonons) traveling on the background fluid. This speed depends on 
the average dynamic structure (determined by mean, variance, and maybe
higher orders, of density, speed and acceleration on the underlying fluid) 
on the thermal range. That is

\begin{equation}
c = c(<\rho^n>_{th},<{v}^n>_{th},<{a}^n>_{th},<{e}^n>_{th})
\label{soundSpeed}
\end{equation}

This speed is homogeneous in a certain neighbourhood while the distributions 
of $\rho$, $\mathbf{v}$, $\mathbf{a}$ and $\mathbf{e}$ in the neighbourhood 
are homogeneous. This is true at the macroscopic range
but fails at the finer ranges. Even at macroscopic ranges, the local sound 
speed may change due to variations in the structure of the background. 
Superluminal signalling can then be possible if the ether is affected by 
macroscopic phenomena. The Scharnhorst effect, for instance, induces a 
variation of $<\rho>$ at a macroscopic range\footnote{We recall that
$<\rho>\equiv<\rho>_{m}=<\rho>_{th}$}, so affecting the speed of sound.

\subsubsection{Lorentz invariance}
\label{lorentzInvarianceOrigin}

Consider a reference system described by its speed $\mathbf{v_O}$ 
and acceleration $\mathbf{a_O}$ relative to the average speed $<v>_m$ 
and acceleration $<a>_m$ of the underlying ether. Suppose that
we are studying a macroscopic-range phonon. In this approximation we can
separately describe the macroscopic wave and the underlying ether. 


We will take a reference system comoving with a macroscopic wave
representing a fermion, which can travel at speeds which range from $0$ to
$c$ with respect to the ether. We are
interested in finding out which quantities can participate in the
effective Lagrangian $\mathcal{L}_m$ derived from the equation 
(\ref{particleLagrangian}), whose detailed description we have delayed and 
that approximately describes the macroscopic wave disregarding the background 
(incorporating it through constants and modifications of the terms in the
Lagrangian), and how Lorentz invariance (or a whole
effective Minkowskian space-time) emerges from this
approximation.

We start by considering the superposition of the macroscopic wave (the
particle) and the background. The particle distinguishes from the
background in that it has an internal structure in the thermal range, which means that 
in the comoving reference system the integration on times $dt>>T_{th}$ of 
$<\partial_i a_j>_{th}$ differs from zero on different regions of the 
particle in this range, while this does not happen to the background, which
is chaotic. In the effective macroscopic-range Lagrangian $\mathcal{L}_m$, 
this internal relative accelerations are concealed
from the point of view of the simplified macroscopic description of the 
particle (its particle numbers, its kinematics and the forces that
appear between particles).

Now consider the dynamics emerging from the kinematical relation between
the particle and the background. If the particle accelerates relative to
the background, we can decompose the gradients of acceleration in two
terms: the internal particle gradient which comes from the own particle
structure and the gradient coming from the relative acceleration of the
particle to the background. This second term provides a dynamical effect, 
according to the nature of the fundamental Lagrangian, that only disappears 
when the relative acceleration is smeared (i.e., inertia). We will consider 
this further when discussing the second pillar. The dynamics provided by 
the fundamental Lagrangian should provide a mechanism for dissipating the 
relative accelerations. If we consider a gradient in speed now, we can as well
decompose the gradients of speed in internal and background-relative. This
terms does not provide any dynamical effect, according to the Lagrangian.
Then, as the fundamental Lagrangian is invariant under boosts, the
equations for the particle disregarding the background are Lorentz invariant. 
This translates into Lorentz
invariant equations when we consider an effective Lagrangian describing the 
interaction between a particle and the background.

Lorentz invariance is then provided by the nature of the underlying ether
force: it only depends on relative accelerations. Then, effective
macroscopic-range forces as (we assume) electromagnetism, can only depend
on relative accelerations $\mathbf{a_O}$ to the ether (but not on relative 
speeds $\mathbf{v_O}$ to the ether). Of course, initial data, which includes 
typical speeds in the stochastic motion of the fluid ($<v>_{th}$) and 
the sound speed $c$ of macroscopic waves (\ref{soundSpeed}), can enter the 
effective equations. The important thing is none of these speeds depends on 
the relative speed of the reference system to the underlying ether. 

Relative speeds between particles could in general also have dynamical 
effects. This is because the equilibrium values, in terms of the $v$ and
$a$ fields of a particle in the thermal range are perturbed by the presence 
of another particle.
Any change, either in relative position or speed, of this perturbation
has a dynamical effect, as the internal gradients of acceleration of the
particle need to accommodate to the perturbation. Relative position and
speed to the background, on the contrary, do not perturb the equilibrium
of the particle. 

Summarising, the interaction of a stable particle with the background
provides the phenomenology of Special Relativity. On this approximation
a particle can be described kinematically through constants and the
assumption of a background Minkowskian space-time. When we consider
more than one particle it is not possible to keep this simplified
description, as the vortices affect each other. From the point of 
view of effective theories we can still maintain the description of the
particle as if isolated, but the price to pay is the introduction of forces
or interactions between the particles.






These elements are enough to derive Special Relativity~\cite{LIB02} on the 
macroscopic range
for any interaction that derives from the ether force. 
This provides an underlying hypothesis for the Lorentz's approach on
Special Relativity. The space-time of orthodox Special Relativity is 
downgraded to a helping tool with no profound meaning. The underlying space 
is Euclidean and can coexist with an absolute time.







This point of view can be easily understood with a fluid example due to LSV
\cite{LIB02}. They consider a fluid Universe, with hypothetical beings whose 
internal structure is completely mediated by phonon exchange, and whose rulers 
and clocks are likewise held together by phonon exchange, and who are
completely blind to electromagnetism so they cannot not discover the real
underlying electromagnetic basis of their laws. They would discover an
\textit{acoustic Lorentz invariance} with their rulers and clocks
transforming according to the laws of approximate \textit{acoustic
relativity} (Lorentz group with $c$ as the sound speed). With this, motion
becomes undetectable\footnote{As we have seen, this is achieved by forbidding 
friction-related speed effects, which do not appear in the proposed form of 
fundamental equations. The background fluid, which we have called ether due to
the lack of a better word, cannot be detected by its effect on relative speeds,
but only on relative acceleration, which permits dodging the inconsistencies
of the XIXth century ether models.}. The fluid would play the role of an ether
whose presence is masked by the length contraction and time dilation
effects. Thus, we would have a fluid dynamical analogue of special
relativity \`a la Lorentz. Of course the underlying physics (electromagnetism)
is also Lorentz invariant in this example (although with a different sound 
speed), but this is not essential. 

\subsection{The second pillar. Gravity and inertia from an ether force}

Gravity and inertia can also find a place as effective theories on
the macroscopic range.

We believe that inertia is a manifestation of the average ether forces
on accelerating objects. We will try to describe the mechanism on which
this could be based. Along the way, the first steps on an explanation
for WFC, to be completed in~\ref{wfcOrigin} will appear.

To explain gravity, we adhere to the group of theories which derive it
from the quantum (here understood as thermal-range effects) phenomenology, 
like those described in section \ref{emergentGravity} (BVL, HRP), although 
rather than deriving gravity from a particular quantum field, we should derive
it from the long-range properties of the average ether forces. 

\subsubsection{The origin of inertia}
\label{inertiaOrigin}

How does inertia emerge in this framework? Consider phonons
on the macroscopic range (in the SFM picture, at the moment). Acceleration 
generates acceleration in waves with homogeneous spin, considering our 
assumption of a spin-driven ether force. 

The equilibrium state in the macroscopic range, then, must be that with 
no relative acceleration between the waves and the background. When a 
macroscopic wave accelerates respect to the ether in its neighbourhood the 
equilibrium is broken and can be restored by two alternative outcomes: 

\begin{enumerate}
\item The macroscopic-range phonon may manage to drag part of the surrounding 
thermal-range ether background on which it travels (its supporting inertial
system, in fact) along its movement,
and so to cause a WFC. If it can drag enough stuff to
account for the numbers describing a metastable vortex state, then
we get a temporary condensed state that we call
particle. Of course this can only happen when the dragging collapse
has a central point and develops radially. This is the only time 
when particles \textit{exist}, as very condensed wave-structures 
(but not point-like anyway): the rest of the time they behave 
as purer (in the sense of not so non-linear) waves. This point only 
makes sense within the SFM.

\item The ether background, over which the wave travels, manages to stop 
the accelerating wave. This is what we call inertia: resistance to relative
accelerations. It is useful to see it from the perspective
of the accelerating wave: it is initially stopped, and the
rest of the Universe accelerates homogeneously, so the wave
is at the end dragged and accelerated to keep the same
acceleration as the rest of the (local) Universe. In this case
the acceleration is linear rather than radial. This second outcome is
feasible even outside the SFM.
\end{enumerate}

How does this hypothesis compare with the two approaches to inertia
mentioned in section~\ref{secondpillar}? It follows the basics of the HPR 
hypothesis. A local force, related to the void (in the sense of
subsection~\ref{inertia}), generates inertia, which is a variant of the HPR 
model. The local void (the ether in the nomenclature of this paper) is
in the end provided by the surrounding Universe (the far stars), or
at least is in equilibrium with it, so the fundamental part in the view of
the Mach School is preserved\footnote{What is not preserved is Einstein's 
implementation.}. For instance, if we decrease the density of
matter in the Universe then the equilibrium at a fixed time is
restored by decreasing the density of the ether $<\rho>$ (as in the SED model),
which in turn decreases the intensity of inertial forces. Inertia is therefore
provided by the far stars as in Mach's view in some sense. However, 
the interaction is not instantaneous, but mediated by a \textit{field} in 
equilibrium with the far stars: the ether field. With this, the need for a 
mechanism explaining the instantaneous interaction of matter here with the far 
stars there vanishes.
Also, inertia cannot be derived from gravity, for it is a rather 
fundamental mechanism. To explain it, no reference to gravity has been made.
And what is more (see next subsection) we think the SFM can explain
gravity as an inertial force.

It seems, summarising, that the ether force has a strong link with inertia. 
Inertia is a stochastic short range manifestation of this fundamental force. 
The Principle of Equivalence forces us to attribute the origin of weight to 
this same force, as commented in section~\ref{secondpillar}. Further analysis,
however, is needed to explain geometrodynamics.

\subsubsection{The origin of gravity}
\label{classicalGravity}

As we expect the spins in the cosmic void background to be
homogeneously distributed, as they are in matter, big concentrations
of matter should feel ether forces as independent of spin (within the SFM;
without the extra SFM assumptions we do not need this consideration in the 
classical approach). The mass
(inertia) of a simple wave is explained regarding to the ether forces
that appear when it is accelerated respect to the ether. The inertial
mass of macroscopic objects can then be explained as an average spin
force in the SFM, and generally as an average ether force in the classical
approach.

In the approach in subsection~\ref{lorentzInvarianceOrigin} that permits the 
derivation of Special Relativity, we assumed no influence of the matter onto 
the ether term. This is true in a limit case where matter can be considered 
in the sense of test particles on the background. The direct influence of the 
background on matter, that is inertia, is in fact part of the picture that
explains the emergence of Special Relativity. But we still can consider the
effect of the matter term on the background. Generally, the matter term can
modify the surrounding background. This could imply, for instance, a
modification of the average density or other parameters. This is reflected 
on a change on the basic parameters defining the background ($c$ and $\hbar$ 
in particular, as functions of $<\rho^n>$, $<{v}^n>$, $<{a}^n>$, and 
$<{e}^n>$). 
From this point of view we can then forget the source of the 
perturbation and describe the resulting scenario purely on the basis of the 
new statistical parameters which describe the ether. These modified properties 
of the ether will 
influence any test particle in the same way, which raises the opportunity of 
describing the picture in geometrical terms. A connection should be provided 
for the basic parameters of the geometrical description (the metric) and the 
statistical parameters describing the ether background. For instance:

\begin{equation}
g_{ij}(x,t) = g_{ij}(c(x,t),\hbar(x,t),...)
\end{equation}

plus gauge conditions to define the coordinate system\footnote{
A natural choice of gauge conditions would be to take a null shift 
($g_{i0}(x_i)=0$, with the spatial coordinates comoving with the ether) and 
$dx^0=dt$ with $t$ representing the absolute time (so that $g_{00}$ reflects 
the slowing of the physical process due to the kinematical changes of the 
underlying ether).}.

This approach can be considered in the line of Unruh's method~\cite{UNR95} on 
deriving the metric from the Lagrangian of the matter fields. It is relevant 
to consider the work by Barcel\'o and Visser~\cite{BAR01}, which shows how 
long-range effects of reasonable quantum fields can induce a geometrodynamics.
Here, however, we are a step below QFT.

In subsection~\ref{gravityZPF} we will analyze how the incompatibilities 
between General Relativity and Quantum Theory can be approached. 

We will try now to precise further how matter modifies the surrounding ether.
This is only valid in the SFM.
In our discussion on the fourth pillar we described fermions and some bosons 
as divergenceless vortices of the fundamental fluid. We will analyse the 
asymptotic effect of a large aggregation of such vortices at large distances.

Notice first that a static vortex, independently of its particular structure,
is reflected in an acceleration field with certain characteristics. A
divergenceless vortex needs that the average acceleration at a certain
distance to the vortex is radial and directed inwards. Of course, at certain
angles the acceleration can be directed outwards, but not on average.
Generally, the vortices are described by a symmetry axis which determines
the orientation of the acceleration field around the vortex. If we consider
a conglomerate of such vortices so that the axis is randomly oriented in
space, then the acceleration field at a large distance from the conglomerate
is spherically symmetric and points inwards.

We now consider how this affects the structure of the stochastic ether
background at a large distance from the conglomerate. Away from any body, the
average acceleration and speed of the ether is null everywhere in the comoving
global reference system. In the
neighbourhood of a conglomerate of vortices, however, there is a mean radial
acceleration towards the body. This does not mean that the ether falls
down into the body: simply it needs to locally rotate around the body.
Consider a small volume of the ether away from the body.
Let us say that the plane perpendicular to the radial acceleration is 
locally described by the $xy$ coordinates, while $z$ represents the radial 
axis.
The module of the thermal speed of the ether noise 
($<|v|>_{th} \equiv <\sqrt{v^2}>_{th} \ne 0$) will increase in the $xy$
plane, so that we can describe the motion of the ether as a random noise
plus a constant radial acceleration towards the body.

If $<|v|>_{th}$ and $<|a|>_{th}$ represent the average module of the speed and 
acceleration of the ether fluid away from bodies in the ether comoving
reference system, $a_g$ is the average 
radial acceleration at a distance $R$ of the body, and $v_g$ is the average 
increase in tangential speed, then in the vicinity of a body we have that
in the comoving reference system

\begin{eqnarray}
 <|v_z|>_{th} & =& <|v|>_{th},\\
 <|v_{xy}|>_{th} & =& <|v|>_{th} + v_g,\\
 <|a_z|>_{th}& =& <|a|>_{th},\\
 <|a_{xy}|>_{th}& =& <|a|>_{th} - a_g,\\
 a_g& =& {{v_g^2}\over{R}}  
\end{eqnarray}

The overall effect is that although there is not a macroscopic axis of
rotation for the ether around the body, microscopically (in the thermal range)
the ether is orbiting the body. The ether accelerates towards the body although
there is no net flux of fluid around any spherical surface centered on the 
body.

Now we concentrate on the effect of this phenomenon on a test particle.
We know that inertial forces tend to stick test particles to the ether
in terms of acceleration, so the test particle \textit{feels} the acceleration
field $a_g$ generated by the body. If $a_O$ is the acceleration of the 
body, then $a_O=<a>_{th}-a_g$. In the comoving frame (of the ether at a great
distance from the body) $a_O=-a_g$. As the test particle is 
characterised by a stable structure, this acceleration cannot be manifested 
microscopically as rotations around different axis, but actually falling into
the body or rotating about a well-defined axis. 
This attraction is independent of the characteristics of the test particle, 
so it can be defined in terms of geodesics, or geometry.

Once we have geodesics, if we add the requirement that the acceleration
is of the Newtonian type in classical regimes, that is $v_g$ proportional
to $\sqrt{M}$ (where $M$ represents the inertial mass of the body), we 
automatically have General Relativity as an emergent phenomenon. This condition
poses a restriction on the particular choices we can take on 
(\ref{subthermalLagrangian}), as it partially defines the long distance 
structure of the stable vortices that are solutions of 
(\ref{subthermalLagrangian}).

In conclusion, gravity (weight) is an inertial reaction force caused by the 
perturbation that a massive body exerts on its neighboring ether. The
gravitational mass of a body is the same as its inertial mass because
gravity is just an inertial force. Notice that the geometrodynamical
approximation is valid in the macroscopic range, but not in the thermal one.
This releases us from the need of a Quantum Gravity.

\subsection{The third pillar. Indeterminism from noise, 
collapse from ether force}

The models of inertia and Special Relativity described above can be fitted in 
the classical approach in general. For the explanation of General 
Relativity we have needed specific features of the SFM. The same happens 
when modelling Quantum Theory. We lean on two phenomena: Pearle's random 
choosers and WFC. The former phenomenon 
can be generally described within the classical approach. The latter, however, 
needs the assumptions we defined as the SFM.

\label{qftOrigin}

With this, our framework may provide a basis for Quantum Theory as an effective
theory on the thermal range. Let us explain how to model the apparent 
fundamental indeterminism and the collapse process.

\subsubsection{The origin of indeterminism}

At the thermal range the local values of density, speed and acceleration 
of the ether are randomly distributed following
a noise pattern, in a sort of Brownian motion. As explained in section
\ref{thirdPillar}, stochastic classical interactions with a background 
noise (Pearle's chooser) can reproduce the probabilistic features of QFT. 
The basic problem of providing a source for the noise is 
trivially solved here. 

Notice that being the ether our model of ZPF, Quantum Theory is derived
from the ZPF, in opposition to the orthodox view. This supports in part the 
SED vision (see section \ref{holes}) and the HPR notions on the ZPF.

The question of a dynamical model for collapse needs a more detailed
analysis. 

\subsubsection{The origin of the Wave Function Collapse}
\label{wfcOrigin}

Can we complete this model with an analogue for the WFC? We have
supposed that ether forces originate it when a variation of
acceleration happens. This initial seed gets amplified by the
characteristics of the fundamental force by unleashing a runaway solution.
But then, what causes the seminal gradient in acceleration on the wave? 
There can be just one answer: the detector as perturbed by the surrounding
noise. We find here an alternative 
explanation to the known fact in QFT that the measure perturbs the 
system (that is the particle):
that is right, since the measure is the only way to get the particle, as in
Jordan's view. Particles only exist when they are measured  (we refer to
\textit{real} particles; for a discussion on \textit{virtual} particles 
see the end of the subsection~\ref{virtualParticles}). In fact, the 
measure builds the particle when the detector causes the component of
the inertial system sustaining the particle to collapse towards the 
detector itself. 

To get a measure we need that some part of the detector (an atom, an
electron) starts to interact with the external wave. Notice that the
part of the detector is also a wave in the fluid, so we should assume
that there is a strong non-linear wave interaction that causes the
initial acceleration. The end of the process is the destruction of the
relatively stable initial state of the detector by the collapse of the
external wave upon it, and the formation of a new metastable state.

The WFC would be a phenomenon parallel to the gravity-induced dragging of 
inertial frames, although in the WFC-dragging the spin would play a major 
role. We will complete the explanation when we discuss how the EPR paradox 
is to be approached from our point of view.

\subsection{Aspects of the dynamics of ether}

Assuming the explanations for the origin of the four pillars we have just
developed, there are several scenarios and phenomena that need to be
analysed under this new prism. In this subsection we tackle some of them.

\subsubsection{EPR and collapse-related issues}
\label{explainepr}
Let us recall EPR. We will try to complete the explanation of subsection
\ref{wfcOrigin} on the dynamics of WFC and its relationship with inertia.

Suppose we have emitted two electrons in the spin singlet state. The total
spin of the system is zero and we have two detectors that are prepared
to detect electrons with opposed spin projections. In standard Quantum Theory,
one says that both electrons share a wave function that is distributed in 
space, the total spin of which is zero. When one of the electrons interacts 
with the detector it collapses within a certain spin: its (part of the)
wave function suffers a WFC. The other electron can be meters away and
collapse almost immediately into the second detector, also by WFC. The
spin of the second electron is what it needs to be in order to conserve the
total spin of the original system. However, in many situations there
is no time for a non-superluminal signal to travel from the place of
the first collapse to the second to inform about the spin state of the
first electron. 

Most attempts in modelling this system in a deterministic way adopt
the point-like paradigm: the electrons are real and concentrated on a
small volume or a point, so either there are some variables that a
priori determine the outcome of the spin measure or a signal is send
in a non-local way. While we design the models to respect the
point-like paradigm and the principles of relativity, they are shown to fail 
to explain experiments (by Bell's theorem), unless non-local influences from
detectors to distant particles are introduced. It would be very 
complex to provide an underlying deterministic mechanism for such an influence
on point-like particles. In our opinion, the burden for the hidden variables 
approaches is that they have tried to attack the third pillar while leaving 
the first (Special Relativity) or fourth (point-like particles) untouched.

Let us explore the situation in the frame of our alternative approach. We have 
recovered determinism, and superseded the point-like paradigm and the speed of
light in vacuum as a bound speed, as we have described them both as emergent 
phenomena. Imagine that when two electrons are emitted they
are just a system described by a classical wave in the strict sense
(so we do not assume the probabilistic axiom of standard QFT), without
any particle property. We are then not allowed to talk about two
electrons individually, but just about a wave bearing the charge (-2)
and spin (0) equivalent to two electrons\footnote{This is in fact the
Jordan-Schr\"odinger point of view.}. When emitted, the macroscopic
wave expands and mixes with the ether and, at some time, part of this wave 
interacts with one of the detectors. What happens at that moment is a very 
remarkable fact, for it is highly non-linear: a WFC. We do not want to consider
it axiomatically but dynamically. The detector can only interact with
a wave with certain spin (say $+1/2$), and so it does: it selects, from the 
wave surrounding it, the spin component that matches its own configuration. 
This surrounding wave must be attracted by some interaction with the 
detector, if we ever want to explain the phenomenon dynamically. So let us 
recall that a force is acting there (which derives from the SFM). Clearly a 
force such this cannot be directly derived from 
electromagnetism or the nuclear forces, as it is inertia-like (that is, 
acceleration-driven) due to the characteristics of the collapse process. 
This fact ranks collapse as an interaction closely derived from the 
fundamental ether force (which we have already used to explain inertia and
gravity), as opposite, for instance, to electromagnetism, which does not have 
traces of acceleration dynamics and then needs to be considered as less 
fundamental. The ether force makes the (spin $+1/2$ projection of the)
surrounding wave collapse towards the detector. Remember that we have
postulated the ether force to be spin-selective. 

However, in order to
get the equivalent properties to one electron concentrated at the
detector, the component of the wave with the right $+1/2$ spin has to be
summoned from all over the entire wave, even from the part close to
the second detector. Let us explain later what these facts imply, and
assume by now we can explain them. The result of the process would be
the following: by ether force, detector one has attracted in a superluminal
fashion (driven by a runaway solution) the part of
the wave that is equivalent to one electron with a certain spin ($+1/2$,
in this case). The rest of the wave remains distributed about the
original region, and has the properties of an electron with the right
spin ($-1/2$) to match the second detector. So now, the second detector
can repeat the process and attract the remaining spin $-1/2$ wave (or it
can fail to do that, so we would say that the second electron escaped
undetected). Up to now the explanation is quite the same as in Quantum
Theory, except that we take the wave function as a closer representation
of the actual underlying reality
(rather than the point-like electron probabilistically distributed by
the amplitude of the wave), so we ignore probabilistic factors, and
that we try to include a model of the WFC in the picture.

We go back now to the pending issue of how the ether force causes the
WFC of half of the original wave (the half with appropriate spin
$+1/2$). First, as advanced, one can see that the collapse must be 
superluminal. 
We use now here our interpretation of the first pillar. If a force drives it, 
then this force does not apparently respect the concept of $c$ as a bound 
speed. We need to provide a mechanism for accelerating
parts of a wave beyond $c$ in almost zero time (but not exactly zero, if
we keep dynamical). This can be done through a model supporting runaway 
solutions (which~(\ref{subthermalLagrangian}) provides). This acceleration 
needs to be spin-selective: it
has to be null for all components of spin except for the one that
initiated the collapse. And this is in fact the main reason to postulate that
the ether force is spin-selective.

Of course, this picture requires some explanation on the role of the
conservation of energy and momentum. To tackle
the apparently instantaneous collapse we have invoked self-acceleration:
acceleration in any part of the wave should cause acceleration in
nearby regions\footnote{This is one of the reasons for the first order
derivatives in~(\ref{subthermalLagrangian}).}. Only this way we can explain 
that the initial collapse near the detector propagates at an enormous speed 
through the entire wave. We know, though, that this behaviour can induce 
violations of the principle of conservation of energy in the form of runaway 
solutions. Runaway solutions are forbidden in traditional physics. However, 
in a random ether background governed by an acceleration-driven ether force,
the runaway tendency can get controlled (that is, limited in time and behaving
stochastically). Local runaway solutions do
not imply global runaway solutions. The question of the conservation
laws will be commented on section (\ref{conservationLaws}).

Another point of view to take into account is that of inertia.
As the wave has a certain energy, which means certain inertia, then this 
inertia needs to be somehow suspended or modified to allow for the action of 
the ether force with its fantastic accelerations. Notice that we could
dismiss the need for an explanation for the role of inertia, since we
have developed the hypothesis that inertia is an effective interaction in
the macroscopic range (so less fundamental than collapse, in fact), but it 
is instructive to see how we could adapt the theory of inertia to
keep it valid even in collapse-driven situations.

In Special Relativity, the speeds are of course always measured within
an inertial system, within which the speed cannot
usually trespass the limit $c$ (we have seen exceptions to that in 
subsection~\ref{violatingRelativity}, soft violations in a sense). Even in WFC,
and EPR in particular, we may consider that no hard violation of the Principle
of Relativity occurs if the inertial system can be described as collapsing 
in parallel with the electron. In general, any dynamical and superluminal WFC 
would be explainable even without any knowledge of the underlying theory if 
there were a collapse of the inertial system sustaining the collapsing wave. 
The ether force, as described above, is spin-selective, so that the inertial 
systems would also be spin-selective. That is important for the explanation of
EPR: in the first collapse in EPR, the inertial system for the spin
measured in the first detector would collapse towards the
detector. The orthogonal inertial system would remain unaffected, but
would collapse later together with the collapse of the rest of the
wave (the second electron)\footnote{This birefringence of the inertial system 
reflects a property from the fundamental underlying force, and also affects 
gravity in principle, although because of the statistical nature of gravity, 
the effect is irrelevant.}. It is tautological to remark that inertial
systems are determined by inertia, that is, in our view, by the
stochastic background. Therefore, a spin
component of the stochastic background is wiped out by the WFC in the
present hypothesis. This creates the condition for a negative
density of energy\footnote{Which, as discussed in \ref{violatingRelativity}, 
is the stigma of superluminal phenomena. Here we see the reason why.}: that is 
because we usually define the zero of energy as the energy that holds the 
normal vacuum\footnote{This is sensible in macroscopic-range forces, as
inertia and gravity, because these are forces that operate on an idealised
background, which has then to be ignored. In thermal-range phenomena, the
energy does need to be considered, as we do not operate in such an
effective background, but we are describing it together with its 
perturbations.}, which is not void at all, but filled with a 
stochastic background field. 
Consequently, anything that reduces the density of energy of the vacuum 
induces formally a negative energy density; a reduction of inertia, in other
words (like in the Casimir plates).

A final point to discuss is that, in fact, the \textit{suction} does not 
need to extend indefinitely in space. Detectors cause WFC all the time,
even when there is not any electronic stuff around. This is because electrons
are made of the same fluid stuff as the ether. Then the detector can always
steal part of the surrounding ether to create a virtual electronic WFC. 
However, this creates a zone of negative energy around the detector, that 
will tend to be filled again by the stolen fluid.

When we have an excess density provided by the electronic wave, on the other
hand, the zone of negative energy can be compensated by the electronic wave
rather than by the fluid collapsed on the detector. Then the wave of negative
energy advances (faster than $c$) from the detectors outwards, while filled 
by the electronic wave, and finally vanishes as completely cancels with the
positive energy wave representing the electron. This way, the virtual collapse 
becomes real thanks to the excess fluid above the average density which was 
released as an electronic wave\footnote{Notice that this effect does not
provide different predictions from Standard Quantum Theory, as the process
can be described quantically as a chain of interactions between the detector
and a cloud of virtual particles surrounding it and between this cloud and
the real electron.}. The collapse is then
local and almost instantaneous, as it feeds initially from ether surrounding
the detector. The presence of the electron simply avoids the reversal of the
process.

If this framework makes sense, then it explains why the reasons of EPR
against Quantum Theory are unsuccessful. They had in mind a relative 
independence of well-separated objects in space (say $A$ and $B$): an external 
influence on $A$ has no direct influence whatsoever on $B$. But this is false 
in our opinion: an external action on $A$ means a measure on $A$, which means 
a collapse and a modification of all the surrounding ether, including the 
part close to $B$, so $B$ is affected. $A$ influences $B$ through changes in 
the structure of the void, and the influence is superluminal (although 
dynamical and finite in speed). Summarising, it is the ether itself that 
collapses on $A$, and the collapse of the ether creates the particle. 

\subsubsection{Real and virtual particles}
\label{virtualParticles}

Now within this scheme we can also explain virtual particles. As advanced,
the same phenomenon described above can happen spontaneously in the ether,
where some local attractor instability (a forming vortex-like structure)
plays the role of the detector. The main difference is that here there 
is not initially a spare energy density above that of the mean void 
(which is the footprint of matter), so the energy subtracted, for instance, to
create a pair of virtual particle-antiparticle creates a zone of negative 
energy, that is, reduces the intensity of the ether. This system is obviously
unstable, as random fluctuations will tend re-equilibrate the density
of energy, so destroying the newly created particles. 
In fact, the only difference between real
and virtual particles would be that real particles are surrounded with
a saturated thermal bath, while virtual particles are surrounded by a
bath with a comparatively low energy density, which tends to re-absorb the 
particle to reestablish the equilibrium.

The magnitude of such a phenomenon is provided by the Heisenberg relation on
uncertainty. $\hbar$ becomes the measure for the transition from the thermal
range to the macroscopic range: even when energy and momentum are not
strictly conserved in the thermal range, the violations are stochastically
limited in time and space. This permits temporary violations and then all the 
phenomena linked to virtual particles, which implies a violation of energy 
and momentum. We will provide more detail in the next subsection.

\subsubsection{Conservation laws}
\label{conservationLaws}

A major problem with a model including runaway solutions, like ours, is 
the role of the conservation laws (conservation of energy-momentum, in 
particular). These laws need to be downgraded or reinterpreted. 

We have assumed our model is built on an isotropic and homogeneous space.
Regarding the Noether theorem, the model presents invariance over time
displacements and space displacements in the framework of the Euclidean
space and absolute time. That is, the fundamental equation for the dynamics 
of the fluid must incorporate 4 conservation laws linked to these
displacements. Also, 3 additional conservation laws emerge due to
the Euclidean space invariance under space rotations. Hence, the
equivalent concepts of energy, momentum and angular momentum are valid 
at the level of the fundamental dynamics of the fluid. There is no Lorentz 
invariance associated to them and at this level a limiting speed is 
not included in the dynamics. These are, of course, not the standard
conservation laws of the Standard Model. In fact, the conserved
quantities can include variables such as relative acceleration and
its derivatives, but not relative speeds.

To obtain the usual conservation laws we need to consider the
approximate theories on the macroscopic range. In this range the ether
constitutes a background with the properties of spatial homogeneity,
time homogeneity and fulfillment of Special Relativity (as a sound
speed has emerged and Lorentz invariance applies). Effective 
theories in this framework can be built on a effective Minkowskian space-time 
disregarding the ether background.
For these effective theories, effective conservation laws will emerge
for energy-momentum in the standard relativistic form.
As this laws are based on the homogeneity of the Minkowskian space-time,
they are accurate while the Minkowskian space-time approximation is valid,
that is, at the ranges where the ether background is homogeneous and
isotropic (this is our definition of the macroscopic range). These are the 
standard conservation laws.

The standard laws are violated, in particular, at the thermal range,
which is the playground of Quantum Mechanics. At this range, the effective
Minkowskian space-time is not longer homogeneous. The ether noise shows up
at this range, allowing for local violations of energy-momentum. A
stochastic local increase of the ether density would appear to 
increase the classical energy through a violation of its conservation
law. The Heisenberg relation gives the order of magnitude of this phenomenon.
That means that local violations of the energy conditions happen all the
time. However, we can set a weaker averaged energy condition so that
energy conditions are approximately respected when appropriate spatial
and temporal means are considered. The ether dynamics tends to stochastically
compensate deficits or surpluses of energy and keep the equilibrium. This
provides a basis for the \textit{Quantum Interest Conjecture} of Ford and 
Roman\footnote{See, for instance the review by Barcel\'o and Visser
~\cite{BAR02c}.}.

An analogue for that would be a probe particle affected by the
Brownian motion of a surrounding fluid. In the \textit{classical limit}
(big particle), one can simplify the model by representing the
noisy background fluid by an homogeneous framework and then obtain an
effective theory for the dynamics of the particle, including conservation
laws derived from the homogeneity of the background fluid (in
particular, the particle moves on a straight line). This approximation
fails when the mass of the particle is not much bigger than the
mass of the particles that constitute the fluid. In this case, the
simplified background is not homogeneous anymore, and the probe particle
can show violations of the effective conservation laws (for instance,
its movement can depart from a perfect straight line, which is a
violation of the momentum conservation law of the simplified model).

Not only the thermal range shows violations. At classical ranges violations 
are not ruled out by the model. In fact,
those violations are currently accepted~\cite{BAR02c}.

At the cosmological range, the ether fluid is not homogeneous: is
is spherical and it is in expansion. Then,
violations of energy can appear: the accelerated expansion of the
observable Universe is associated with a dark energy term, as if
classical energy was being constantly introduced all over space. This
gives a reason for the violation of the Strong Energy Condition (SEC).

Regarding the form of the conservation laws, 
the effective Lagrangian for standard theories includes masses
and relative speeds of fields or particles. Inertial masses are 
an effective concept, derived from the short-range ether force, 
as discussed in subsection~\ref{inertiaOrigin}. The standard 
relativistic conservation laws are dependent on these quantities. Kinetic 
energy, for instance, is a concept dependent on the support of an underlying 
effective inertial system, without which it has no sense.

The fundamental conservation laws, on the other hand, cannot depend
neither on inertial mass, since it is a derived concept, nor on
relative speed (as a consequence of Michelson-Morley experiments, or from
our point of view, from the restriction that speed does not appear in the 
fundamental Lagrangian (\ref{subthermalLagrangian})). They can only depend on 
the quantities that enter in the Lagrangian ruling the ether dynamics. We 
have assumed in section~\ref{assumptions} that these quantities are the 
density of the ether fluid $\rho$, and the gradients in acceleration 
$\partial_i a_j$. Then the ether conservation 
laws should be formulated in terms of ether density rather than 
inertial mass, and the relative acceleration, rather than relative speed.  

\subsubsection{Gravity and the ZPF}
\label{gravityZPF}

Regarding our description of gravity on subsection~\ref{classicalGravity}, we 
note that the geometrodynamical approximation is valid while we can consider 
that

\begin{itemize}

\item the interaction of two (or more) masses is not direct (meaning 
quantum-like, or thermal, interactions), but indirect through general (only 
affecting general parameters describing the ether) mutual modification of 
the underlying ether.

\item the neighbouring ether can be considered as an homogeneous kinematical 
background whose detailed description we can override.

\end{itemize}

On this basis we have to note that the ZPF is a part of the kinematical 
underlying description of the background. 
Gravity, on the other hand is an approximation to one of the influences of 
matter on this term. It is not meaningful, then, to consider the background 
as one of the roots of gravity. Gravity is produced by the matter phenomena 
which represent an excess of energy over that of the background. The 
background itself is to be seen as homogeneous and not containing energy
(which is the perspective in the macroscopic range: the ether has energy
from the thermal-range approximation, but not in the macroscopic-range one), 
from the gravitational point of view. Then the void has no weight in the
macroscopic range. The quantum 
description, which needs to consider the energy of the ZPF is more 
fundamental than gravity, in the sense that does not consider the ether as 
a pure kinematical background, but depends on its dynamics.

The General Relativity and Quantum Theory representations are on different 
levels of approximation to the fundamental theory, and they represent 
different regimes. Their conflicts only reflect that. 

\subsubsection{Cosmic evolution}
\label{cosmicEvolution}
A dynamical ether governed by the ether force could explain some of
the cosmological problems of the Standard Model. We mentioned in section
\ref{holes} that the runaway force accelerating the expansion of the 
observable Universe needs some explanation based on the nature of the quantum 
vacuum. In our suggested framework, runaway solutions are implicit. At a
macroscopic range the runaway behaviours cancel out statistically: we
choose the origin of coordinates as $<\mathbf{a}>=0$ as there is no way to 
select a preferred direction;
also we recalled earlier than on the macroscopic range $<\mathbf{a}^2>_m=0$
(compared with cosmological time scales).
This is not true at the smaller thermal range, where runaway dynamics explain 
the WFC, for instance. At a cosmological range, there is an asymmetry along the
radial direction (the expansion of the Universe), so this direction
can present a global runaway solution (its exact nature would depend
on the equations governing the ether force and the initial conditions
of the observable Universe at the Big Bang).

This corresponds with the cosmological constant (if placed on the left in
Einstein equations, as in Einstein's original view\footnote{Einstein's hope
was in fact to eventually move everything to the left term and explain nature 
in terms of geometry.}) or quintessence (if placed 
on the right, as required in the inflaton picture). In fact it is difficult 
to decide a position since dark energy is neither geometry nor matter, but a 
direct result of the more fundamental underlying theory, of which both 
Riemannian geometry
and matter are simplifications in certain regimes. Its dynamics must be 
directly traced to the fundamental Lagrangian~(\ref{subthermalLagrangian}). In 
fact, the dynamics of the cosmological constant, if they can be settled 
experimentally\footnote{There are reasons to believe they will in the near 
future~\cite{PEB02}.} should provide important clues on the real form 
of~(\ref{subthermalLagrangian}).

We can extract some consequences to the dynamics of the ether along the
expansion of the observable Universe.
As we have assigned a conservation law to the amount of ether (or
in general to the cosmic fluid, in ether or matter form), so it is
not created or destroyed, then the ether must thin down with the
expansion of the Universe. This thinning can be traced as the reason
for the diminishing value of the fine structure constant as well as
a possible reason for the anomalous Pioneer acceleration~\cite{RAN02} (this 
was the third problem reported in the void-related unsolved issues in the 
Standard Model).

\subsubsection{Beyond the observable Universe}
\label{cosmos}
On an even more speculative fashion we could imagine the cosmological
scenarios where our observable Universe fits in. If we are just
seeing an expanding blob of fluid, can be something else beyond its
limits? Certainly one can imagine an spatially infinite Euclidean
space (or a finite 3-dimensional surface of a 4-sphere, for instance, if
we prefer a closed model) where Universes like ours explode and mix with 
each other while governed by the ether force. Let us call this broader space 
\textit{Cosmos}, to distinguish it from our observable Universe.

In this scenario, our observable Universe would have been originated
by a Big Crunch, similar in a sense to a WFC at a cosmic range, of part of the 
fluid which fills the Cosmos (this would be natural if we would assume 
invariance of scale in the equation for the ether force). The following 
Big Bang would have created the present structure of our observable Universe, 
which would be expanding, driven by a runaway solution, against a relative void
outside. This outside void would be a much thinner version of the
ether that fills our observable Universe.

The probable fate of the observable Universe in this scenario would be to get 
mixed again with the rest of the Cosmic fluid, like a Supernova sending its 
matter to mix with the galactic clouds of gas, which eventually collapse in
different stars (Big Crunches in our case) in an infinite cycle. Notice that
there is not any singularity, just very dense concentrations of fluid.
The only condition needed to warrant this infinite cyclical behaviour, apart
from appropriate boundary conditions for the Cosmos, 
is that (some) Big Crunches should be able to reverse complex atomic structures
and return the matter to a plasma state like the one that existed in
the first minutes of our Universe. We know that this is true over a certain
density threshold.    

This scenario would also avoid a foreseeable criticism to our model: how do the
departures from cosmological homogeneity emerge in a theory which is not
fundamentally indeterministic? In the traditional inflation framework, the
quantum indeterminacy is invoked to provide the large-scale inhomogeneities.
In a deterministic theory we cannot start with a perfect spherical model, since
even if the model is unstable under small perturbations, a seed must be
nevertheless supplied. The Big Bang must have then started from a relatively
inhomogeneous collapsed state. Most of the inhomogeneity would have been swept
by inflation. The overall acceleration and angular speed, relative to the 
neighbouring fluid, of the blob which gave birth to our Universe is quite 
irrelevant, since inertial effects on the blob depend on the density of the 
fluid around the blob, which is supposed to be very tiny (compared to the 
background ether inside the blob) due to the dragging effect of the Big Crunch 
runaway collapse.  

Notice also that within this picture the initial state of the observable 
Universe is not deductible from~(\ref{subthermalLagrangian}), because $<\rho>$,
$<\mathbf{v}>$ and $<\mathbf{a}>$, in particular, are the result of the 
previous collapse, and the Big Crunches are as stochastic as any WFC. 
Fundamental parameters in 
effective theories should be traced back to this initial conditions plus any 
fundamental constant appearing in (\ref{subthermalLagrangian}). In particular, 
the exact values of $c$, $\hbar$ and $G$ probably do not have any particular 
meaning\footnote{Other than the restrictions based on the anthropic principle.
Presumably many Big Bangs do not have the right initial conditions so that 
life can emerge.}. As Ellis remarks~\cite{ELL01}, \textit{one runs into
major problems in distinguishing boundary conditions from physical laws.
What appears to be an inviolable physical law may just be a consequence
of the particular boundary conditions that happen to hold in this particular
Universe}.

\section{Conclusions}
\label{secConclusions}

In this section we summarize the assumptions we have made during our 
discussion of the classical Euclidean approach and the SFM. Then we also
summarize the main consequences. We then compare the complexities of this 
approach versus the orthodox one (the Standard Model). Considering all this
we finally provide a (subjective) answer to the entitling question of this
paper.

\subsection{Assumptions}
\label{ConclAssumptions}

These are the assumption we have made for constructing the alternative
framework (in parentheses the corresponding assumption under the Standard
Model (SM)):

\subsubsection{Principles:}

\begin{itemize}

\item The fundamental space is Euclidean, homogeneous and isotropic, and 
the time is absolute (SM: the fundamental space-time is Riemannian)

\item The fundamental Lagrangian does not depend on relative speeds
(SM: Lorentz invariance as a given metaprinciple)

\end{itemize}

\subsubsection{Initial data and conditions on the equations:}

\begin{itemize}

\item The space is filled with a fluid in a turbulent thermal state 
(SM: The fundamental laws are indeterministic) 

\item There are stable and metastable vortices as solutions (SM: Point-like
particles exist)

\item Spin is a property that characterises vortex solutions (SM: Spin is an
internal property of particles)

\item The forces between stable and metastable structures are spin-dependent
(SM: The postulate of the quantum collapse)

\item Inertial mass is a property derived by the interaction of vortex
solutions with the thermal background (SM: Inertial mass is an internal 
property of particles)

\item The average increase of tangential speed of the ether around a vortex
is proportional to the square root of the inertial mass of this vortex (SM: 
The Einstein equations)

\item The runaway solutions implicit in the fundamental Lagrangian can
be implosive (SM: The postulate of the quantum collapse)

\item At macroscopic ranges the thermal background is homogeneous and
isotropic (SM: The energy-momentum classical conservation laws) 

\item The runaway solutions implicit in the fundamental Lagrangian can
be explosive (SM: Dark energy and inflation)

\end{itemize}

\subsubsection{Pending issues:}

\begin{itemize}

\item All phenomena are supposed to be derived from the dynamics of a fluid 
(SM: A grand
unification is supposed for the Electroweak and Chromodynamic forces.
Gravity should be included. The particles are treated apart)

\item The fauna of particles should emerge from the characterisation of
these stable solutions (SM: A mathematical structure to be discovered is
expected to provide a reason for the known hierarchies of fundamental
particles.)

\item $c$ is the sound speed of waves on the thermal background and 
can in principle be obtained from other data (SM: $c$ is a 
fundamental constant)

\end{itemize}

So a model for (\ref{subthermalLagrangian}) will need to match all
these conditions. Is it possible? Is it reasonable? Clearly the answer to that
questions still needs a lot of effort, but in our opinion the picture does 
not look as daunting as the string approach.

Notice, furthermore, that if a model can be found that meets all these
criteria, then most of the assumptions become predictions, which
results in an axiomatic system much simpler than the Standard Model.

\subsection{Consequences}

Now assuming such a model can be built, there are
some consequences of a classical Euclidean TOE and of the 
SFM in terms of the qualitative predictions on the structure of particles 
and space, Quantum Theory, Special Relativity, General Relativity, cosmology, 
and the four standard forces.

\subsubsection{Basics}

\begin{itemize}

\item It explains particles as short-lived metastable states that
  result from WFC.

\item Virtual particles have a natural explanation. They only differ from
real particles in the parameters describing the surrounding ether.

\item Energy and momentum are not conserved.

\end{itemize}

\subsubsection{Quantum Theory}

\begin{itemize}

\item It explains the quantum void as a stochastic background in the
  sense of SED. 

\item As in SED, quantum uncertainty emerges from the stochastic
  nature of the background void. This eliminates all the paradoxes related
to the indeterminism.

\item The Feynman-Fivel-Pearle assumption of gravity as the source of 
WFC~\cite{FEY95, FIV97} is partially justified in some way: the ether force, 
of which gravity is a long-range manifestation, is the origin of the dragging
and collapse of the inertial system on which the wave travels.

\item The typical energies
$dE$ that can be subtracted from the void depend on the typical time of
return $dt$ that we consider. This is related to the thermodynamic
properties of the void, depending on its density (of order $h$,
according to SED).

\item All quantum events, real and virtual, are driven by runaway 
solutions.

\item WFC is explained dynamically as a superluminal phenomenon.

\item One of the main characteristics of General Relativity is that 
acceleration causes acceleration or put in another words, acceleration 
drags the inertial systems. That is the result that led Einstein to General 
Relativity in the first place. This behaviour is inherited from the ether 
force. The WFC would be a phenomenon parallel to the gravity-induced dragging
of inertial systems, although in the WFC-dragging the spin would play a major
role.

\item The quantum vacuum does not gravitate, as gravity operates in a
simplified background that includes the ZPF. In particular, gravity is an
inertial force, an inertia does not exist without a supporting inertial
system, which is the ZPF itself.

\end{itemize}

\subsubsection{Special Relativity}

\begin{itemize}

\item Lorentz invariance is an emergent phenomenon

\item The speed of light is not a fundamental constant but the sound
speed of phonons traveling on the ether.

\end{itemize}

\subsubsection{General Relativity and inertia}

\begin{itemize}

\item It explains inertia as a stochastic interaction with the void,
  in the limit of a homogeneous background.

\item Mach's principle (Mach0\footnote{See subsection~\ref{inertia}.}) is 
explained in the 
Spin Force model. From several points of view, the SFM can be considered 
more Machian than General Relativity. For instance, the Mach2 principle reads
\textit{An isolated body in otherwise empty space has no inertia}
\cite{BON96}. Both Newtonian and Einsteinian gravity fail at satisfying
this principle. The SFM, on the other hand, predicts this to be true
(this is relevant even when the direct experiment is probably impossible).
This is because inertia is an effect of the ether in equilibrium with
matter. A void space implies no matter and no ether, and then no inertia.
A piece of matter that could\footnote{It could not escape in fact: it would
evaporate in a sort of micro-Big Bang if it is extracted from the ether
bath.} somehow escape our 
Universe could have any acceleration relatively to other neighbouring 
Universes while travelling on really void space.

Mach3 principle, \textit{Local inertial frames are affected by the cosmic
motion and distribution of matter}, is fulfilled by General Relativity
and the SFM.

Mach7 principle, \textit{If you take away all matter, there is no more
space} is false in General Relativity and in Newtonian gravity. In the
SFM is true, however, if space is understood in the sense of the effective
Minkowskian space provided by the background ether. The explanation is
the same as in Mach2.

From this small survey we can conclude that the SFM is more Machian than any 
other theory, including General Relativity. Of course, this is just a 
notational observation. Scientifically, it is only important whether the 
experimental principle Mach0 is predicted or not.

\item It explains gravity as a long distance inertial effect of the model. In
  this limit, it reproduces General Relativity.

\item Gravity does not need to be quantised.

\end{itemize}

\subsubsection{Cosmology}

\begin{itemize}

\item Cosmological runaway solutions are permitted, which includes inflation
and dark energy.

\item Our observable Universe is enclosed in a wider Cosmos.

\end{itemize}

\subsubsection{The four forces}
\label{fourforces}

In the classical approach the four interactions have a different
relationship from that in the unification extensions of the Standard Model.
We have seen than the geometrodynamics of the Standard Model is replaced with 
three different interactions: an Euclidean gravity in the macroscopic range, 
spin-driven inertia (encompassing quantum collapse, in the thermal range) and 
quintessence (in the cosmological range), all of them deriving
from the fundamental ether force in certain limits.

The role of the three remaining interactions of the Standard Model, that is
Electromagnetism, the Weak force and the Strong force, is more opened. We 
should note that once a model provides spin and Lorentz invariance, then 
certain restrictions appear on the field equations. Spin 1 bosons follow 
Maxwell or Proca equations (also, spin 1/2 fermions are bound to Dirac 
equations)\cite{RYD85}. Then, simple bosons (unstructured phonons) on the 
Spin Force Model, if they exist, are forced to follow Maxwell equations in the
macroscopic range. 
Electric charge, then, seems to be linked to the capability of certain fermion
vortex fields at emitting and absorbing unstructured phonons.
  
We need to invoke complex bosons, with an internal vortex structure (and 
then massive) to account for the nuclear forces. How many of these bosons
there are, if any, is hard to tell without a detailed study of a particular
form of (\ref{subthermalLagrangian}). But some qualitative consequences can
be sorted. First, the Electroweak unification looks quite unnatural in this
framework. Second, Chromodynamics with massless gluons does not seem 
compatible with the framework. If the Spin Force Model is to accommodate
Chromodynamics, then the gluon should have a small mass.

It is also remarkable, when comparing to the Grand Unification models,
that the Higgs boson does not find any place within the SFM.
The origin of mass, in fact, is originated by the inertial interaction,
which is an effect that lays at the same fundamental level than particles.

\subsection{The complexity of the Standard Model vs the classical Euclidean 
approach}

The Standard Model is based on a set of assumptions underlying the
four pillars (mainly, Lorentz invariance and $c$ invariance, the Principle of
Equivalence, the space-time as a geometry and its geometrodynamics, the
absoluteness of space when acceleration is considered and relativeness
when speed is considered, the quantum collapse, the indeterminacy and
the need of a chooser, the point-like particles, duality and renormalisation,
the populations of particles, its internal characteristics, their 3
non-gravitational interactions). All these assumptions have parallel
assumptions in our proposed framework, as we have detailed in subsection
\ref{ConclAssumptions}. We have seen that in this framework, though, the assumptions
seem less arbitrary, as they fit naturally in the model based on fluid
dynamics.

But apart from these collection of assumptions, we need fixes in the
Standard Model for the internal inconsistencies generated by these 
assumptions, which we have analysed in the previous sections:

\begin{itemize}

\item Special Relativity is threatened by the Scharnhorst effect and 
situations involving negative energies. The minimal cure is to degrade $c$ 
from a fundamental constant to a sound speed, which anyway opens a breech in 
the philosophical basis of Special Relativity.

\item The quantum gravity deadlock

\item The quantum void does not gravitate

\item The measurement axiom

\item The renormalisation anomaly

\item Dark energy

\end{itemize}

In the SFM, on the other hand, these are naturally explained, although such a
conclusion has obviously an influence from subjective considerations. 

\subsection{Is the classical Euclidean approach reasonable?}
\label{conclusions}

Is then the search for a classical Euclidean TOE reasonable? Its 
mathematics could be certainly less analytic than what we are used to in
theoretical physics, as specialists in fluid dynamics know and suffer,
but should Occam's razor rely on this factor or on the complexity of
the axioms? We have always weighted favourably the simplicity of the
equations and their analytical solutions because analytical results were,
until recently, the main source of knowledge on our systems of equations.
But that is a circumstantial fact and attaching ourselves to this 
restriction may have led us to an overwhelming
complexity in the underlying axioms. A high price to pay for analytical
simplicity, indeed. Today, however, in the age of computing, 
the complexity of the equations is not a point to consider heavily, as we
can always get numerical solutions. Instead, we should start looking for
a simplicity of assumptions, even when they lead us to non-linear
equations hard to attack analytically. 

In this sense, a classical Euclidean TOE is reasonable, because it can be 
simpler than the Standard Model, whose underlying axioms have entered
a dead path. The interpretational paradigm of the XXth century physics
suffers from a deadlock that already lasts 70 years. The suggested outcomes
that do not deviate from the paradigm are obtrusively complex. 
Paraphrasing Bell, the interpretational paradigm of the XXth century physics
bears the seed of its own destruction.
On the other hand, one can imagine an outcome for the deadlock within the
simple interpretational paradigms of XIXth century physics and fluid mechanics
if one assumes three basic ideas:

\begin{itemize}

\item The Universe is filled with a stochastic fluid background (ether)

\item The interaction of the fluid is mainly driven by its relative
acceleration, that is, we need to assume a dynamics of acceleration
(an ether force)

\item The dynamics is divergenceless and creates vortices

\end{itemize}

Is then a classical TOE reasonable? It certainly seems so.

\appendix
\section{A simple toy model}

One of the most obvious and simplest models for (\ref{subthermalLagrangian})
to try first is

\begin{equation}
\mathcal L(\mathbf{x},t) = \partial_i \mathbf{F}(\mathbf{x},t) 
	\partial_i \mathbf{F}(\mathbf{x},t) - 
	\partial_t \mathbf{F}(\mathbf{x},t)
	\partial_t \mathbf{F}(\mathbf{x},t)
\end{equation} 

where

\begin{equation}
\mathbf{F}(\mathbf{x},t) = \rho(\mathbf{x},t) \mathbf{a}(\mathbf{x},t)
\end{equation} 

and the Einstein summation rule applies (i=1,2,3).
Therefore we have a simple wave equation for the \textit{force} $\mathbf{F}$.

\section{Toroidal vortices as the origin for spin?}
\label{spinOrigin}

If we assimilate fermions to vortices, the simplest assumption is to model
them as toroidal vortices. This might provide a dynamical origin for the
spin. Apart from the standard angular speed axial vector $\mathbf{\omega}$ 
describing
the rotation of the vortex from an external point of view, we need an
additional scalar to account for the internal rotation on the toroidal axis.

Now we postulate that stable vortex configurations are characterised by a
fixed azimuthal-toroidal speed ratio. For spin $1/2$ fundamental fermions,
for instance,

\begin{equation}
{{\omega_\phi}\over{\omega_t}} = 2
\end{equation}

where $\omega_\phi$ represents the angular speed of the vortex around its
azimuthal axis and $\omega_t$ represents the internal toroidal angular
speed.

For such a vortex, the Dirac equation is approximately valid on the 
macroscopic
range. The Dirac equation is derived by the only assumptions of a spin 1/2 
particle plus Lorentz invariance (transformations under the Lorentz group) 
\cite{RYD85}, which is a good approximation on the macroscopic range.

\section{Experimental predictions}

The classical Euclidean approaches, and the SFM in particular,
can be tested through some basic predictions. Here we will list a
collection of predictions from the SFM, although some of them
are traceable to the classical approach. Unfortunately, the predictions
can only be qualitative given the preliminary state of the model. However,
some of the qualitative predictions could be used to falsify the framework.

\subsection{Implications on cosmology}

Currently, cosmological models are confronted with the observations on the
weights of different sources of energy in the framework of the 
Friedmann-Lemaitre model~\cite{PEB02}:

\begin{equation}
\Omega_{M0} + \Omega_{R0} +\Omega_{\Lambda0} +\Omega_{K0} =1 
\label{energyWeights}
\end{equation} 

The qualitative model of section~\ref{alternative} implies that 
$\Omega_{K0}=0$, since the underlying theory is 
Euclidean\footnote{Nonetheless, as mentioned in subsection~\ref{cosmos},
there is no reason to be dogmatic about whether the Cosmos is spatially 
open or closed. The latter option would imply a 
static space curvature, but that would presumably happen on a spatial scale 
much larger than the distances in our observable Universe, so we can consider 
the curvature to be approximately null.}.
Also, the model predicts that $\Omega_{\Lambda0} \ne 0$ in general (see
\ref{cosmicEvolution})

These values are currently the best fits of astronomical observations, so this 
result cannot be considered a prediction. Anyway, more precise observations 
could change that picture and rule out the model proposed in section 
\ref{alternative}.
At present, a $\Omega_{\Lambda0} = 0$ is two or three standard deviations 
from the best fit~\cite{PEB02}. 

The precise values of the weights depend on the form of 
(\ref{subthermalLagrangian}) and on the initial conditions at our
particular Big Bang, so we cannot provide a prediction for the ratios on 
the rest of weights.

\subsection{Increased $c$ in a bath of rotating cylinders}

\subsubsection{The experimental set}
Imagine a long cylinder (the longer the better in order to suppress boundary
effects) containing a bunch of smaller
cylinders. The smaller cylinders should be able to rotate at different 
speeds. The space not occupied by the smaller cylinders should be void, 
both to avoid air turbulence and to permit the measurement of the speed of 
light in vacuum. In particular a tiny space should be left in the middle of 
the device, so that light can travel along it. We will call \textit{container}
the big cylinder and \textit{rotators} the smaller ones. The central corridor 
for the light will be called \textit{cannon}.

At one end of the cannon we place a laser. Light will propagate from the
laser to the other end of the cannon, where a mirror reflects it. An
interference pattern should pop up. The mirror will allow light out to
study the interference.

First we will measure the interference pattern with the rotators
stopped. Then we will accelerate them progressively (in different senses, to
keep the global angular momentum null) while watching any
change in the pattern of interference. 

\subsubsection{Theoretical predictions}

The rotators, when spinning, will create an acceleration field on them,
which will create a spatial gradient on the probability of collapses of
virtual particles. Closer to the rotators, the collapse probability will 
be greater. The thermodynamics will tend to rebalance the density
of the ether by transferring ether from the collapse regions, close to
the internal rotators, to the axis of the cannons. An equilibrium will
be achieved, with a stabilised gradient of ether density: below the
cosmic average on the axis of the cannon and above average in the 
vicinity of the rotators.

The variation in ether density will cause a variation (increase) in speed 
of light, which would be measurable if intense enough.

This effect is similar to gravity, although gravity emerges from the
stochastic microscopical rotation of the ether around a body. Here there
is a macroscopic rotation of the ether.  

\subsection{Masses of fermionic and bosonic fields}

As explained in subsection~\ref{fourforces}, all fermionic fields are bound 
to have a mass within the SFM. Also, all non-trivial bosonic fields 
(Electromagnetism) must be derived from massive bosons. This means, in 
particular, that both neutrinos and gluons should have a mass. 

\subsection{Other differences with Grand Unification models}

The most trivial consequence of the classical approach is that gravity
cannot be unified with the quantum forces, as they have a different
origin. The SFM also misses to find an explanation for a
Electroweak unification (this is commented in the conclusions). We do not
still know whether this is fatal for the SFM. Finally,
the Higgs boson does not find a place in the SFM (see also
the conclusions).

\bibliography{arbona}

\begin{thebibliography}{60}
\expandafter\ifx\csname natexlab\endcsname\relax\def\natexlab#1{#1}\fi
\expandafter\ifx\csname bibnamefont\endcsname\relax
  \def\bibnamefont#1{#1}\fi
\expandafter\ifx\csname bibfnamefont\endcsname\relax
  \def\bibfnamefont#1{#1}\fi
\expandafter\ifx\csname citenamefont\endcsname\relax
  \def\citenamefont#1{#1}\fi
\expandafter\ifx\csname url\endcsname\relax
  \def\url#1{\texttt{#1}}\fi
\expandafter\ifx\csname urlprefix\endcsname\relax\def\urlprefix{URL }\fi
\providecommand{\bibinfo}[2]{#2}
\providecommand{\eprint}[2][]{\url{#2}}

\bibitem[{\citenamefont{Thiemann}()}]{thi01}
\bibinfo{author}{\bibfnamefont{T.}~\bibnamefont{Thiemann}},
  \eprint{gr-qc/0110034}.

\bibitem[{LVR()}]{LVR}
\eprint{http://www.livingreviews.org}.

\bibitem[{\citenamefont{Feynman}(1995)}]{FEY95}
\bibinfo{author}{\bibfnamefont{R.}~\bibnamefont{Feynman}},
  \emph{\bibinfo{title}{Lectures on gravitation}}
  (\bibinfo{publisher}{Addison-Wesley, Reading, Massachusetts},
  \bibinfo{year}{1995}).

\bibitem[{\citenamefont{Fivel}()}]{FIV97}
\bibinfo{author}{\bibfnamefont{D.}~\bibnamefont{Fivel}},
  \eprint{quant-ph/9710042}.

\bibitem[{\citenamefont{Sakharov}(1968)}]{SAK68}
\bibinfo{author}{\bibfnamefont{A.}~\bibnamefont{Sakharov}},
  \bibinfo{journal}{Soviet Physics Doklady} \textbf{\bibinfo{volume}{12}},
  \bibinfo{pages}{1040} (\bibinfo{year}{1968}).

\bibitem[{\citenamefont{Barcel\'o et~al.}()\citenamefont{Barcel\'o, Visser, and
  Liberati}}]{BAR01}
\bibinfo{author}{\bibfnamefont{C.}~\bibnamefont{Barcel\'o}},
  \bibinfo{author}{\bibfnamefont{M.}~\bibnamefont{Visser}}, \bibnamefont{and}
  \bibinfo{author}{\bibfnamefont{S.}~\bibnamefont{Liberati}},
  \emph{\bibinfo{title}{Einstein gravity as an emergent phenomenon?}},
  \eprint{gr-qc/0106002}.

\bibitem[{\citenamefont{Haisch et~al.}(2002)\citenamefont{Haisch, Rueda,
  Nickisch, and Mollere}}]{HAI02}
\bibinfo{author}{\bibfnamefont{B.}~\bibnamefont{Haisch}},
  \bibinfo{author}{\bibfnamefont{A.}~\bibnamefont{Rueda}},
  \bibinfo{author}{\bibfnamefont{L.~J.} \bibnamefont{Nickisch}},
  \bibnamefont{and} \bibinfo{author}{\bibfnamefont{J.}~\bibnamefont{Mollere}}
  (\bibinfo{year}{2002}), \eprint{gr-qc/0209016}.

\bibitem[{\citenamefont{Gibbons}(1998)}]{GIB98}
\bibinfo{author}{\bibfnamefont{G.~W.} \bibnamefont{Gibbons}}
  (\bibinfo{year}{1998}), \eprint{gr-qc/9803065}.

\bibitem[{\citenamefont{Schwarz}(2000)}]{SCH00}
\bibinfo{author}{\bibfnamefont{J.~H.} \bibnamefont{Schwarz}}
  (\bibinfo{year}{2000}), \eprint{hep-ex/0008017}.

\bibitem[{\citenamefont{Feynman}(1965)}]{FEY65}
\bibinfo{author}{\bibfnamefont{R.}~\bibnamefont{Feynman}},
  \emph{\bibinfo{title}{The character of the physical law}}
  (\bibinfo{publisher}{MIT Press, Cambridge, Mass.}, \bibinfo{year}{1965}).

\bibitem[{\citenamefont{Penrose}(1986)}]{PEN86}
\bibinfo{author}{\bibfnamefont{R.}~\bibnamefont{Penrose}},
  \emph{\bibinfo{title}{Quantum concepts in space and time}}
  (\bibinfo{publisher}{Eds. R. Penrose and C.J. Isham, Clarendon Press,
  Oxford}, \bibinfo{year}{1986}), chap. \bibinfo{chapter}{Gravity and State
  Vector Reduction}.

\bibitem[{\citenamefont{Pearle}(1989)}]{PEA89}
\bibinfo{author}{\bibfnamefont{P.}~\bibnamefont{Pearle}},
  \bibinfo{journal}{Phys. Rev. A} \textbf{\bibinfo{volume}{39}},
  \bibinfo{pages}{2277} (\bibinfo{year}{1989}).

\bibitem[{\citenamefont{Einstein et~al.}(1935)\citenamefont{Einstein, Podolsky,
  and Rosen}}]{epr}
\bibinfo{author}{\bibfnamefont{A.}~\bibnamefont{Einstein}},
  \bibinfo{author}{\bibfnamefont{B.}~\bibnamefont{Podolsky}}, \bibnamefont{and}
  \bibinfo{author}{\bibfnamefont{N.}~\bibnamefont{Rosen}},
  \bibinfo{journal}{Phys.\ Rev.} \textbf{\bibinfo{volume}{47}},
  \bibinfo{pages}{777} (\bibinfo{year}{1935}).

\bibitem[{\citenamefont{Ghirardi et~al.}(1986)\citenamefont{Ghirardi, Rimini,
  and Weber}}]{GHI86}
\bibinfo{author}{\bibfnamefont{G.}~\bibnamefont{Ghirardi}},
  \bibinfo{author}{\bibfnamefont{A.}~\bibnamefont{Rimini}}, \bibnamefont{and}
  \bibinfo{author}{\bibfnamefont{T.}~\bibnamefont{Weber}},
  \bibinfo{journal}{Phys. Rev. D} \textbf{\bibinfo{volume}{34}},
  \bibinfo{pages}{470} (\bibinfo{year}{1986}).

\bibitem[{\citenamefont{Ghirardi et~al.}(1990)\citenamefont{Ghirardi, Pearle,
  and Rimini}}]{GHI90}
\bibinfo{author}{\bibfnamefont{G.~C.} \bibnamefont{Ghirardi}},
  \bibinfo{author}{\bibfnamefont{P.}~\bibnamefont{Pearle}}, \bibnamefont{and}
  \bibinfo{author}{\bibfnamefont{A.}~\bibnamefont{Rimini}},
  \bibinfo{journal}{Phys. Rev. A} \textbf{\bibinfo{volume}{42}},
  \bibinfo{pages}{78} (\bibinfo{year}{1990}).

\bibitem[{\citenamefont{Ghirardi et~al.}(1995)\citenamefont{Ghirardi, Grassi,
  and Benatti}}]{GHI95}
\bibinfo{author}{\bibfnamefont{G.}~\bibnamefont{Ghirardi}},
  \bibinfo{author}{\bibfnamefont{R.}~\bibnamefont{Grassi}}, \bibnamefont{and}
  \bibinfo{author}{\bibfnamefont{F.}~\bibnamefont{Benatti}},
  \bibinfo{journal}{Found. Phys.} \textbf{\bibinfo{volume}{25}},
  \bibinfo{pages}{5} (\bibinfo{year}{1995}).

\bibitem[{\citenamefont{Raine}(1981)}]{RAI81}
\bibinfo{author}{\bibfnamefont{D.~J.} \bibnamefont{Raine}},
  \bibinfo{journal}{Reports of Progress in Physics}
  \textbf{\bibinfo{volume}{44}}, \bibinfo{pages}{1151} (\bibinfo{year}{1981}).

\bibitem[{\citenamefont{Woodward and Mahood}(1999)}]{WOO99}
\bibinfo{author}{\bibfnamefont{J.~F.} \bibnamefont{Woodward}} \bibnamefont{and}
  \bibinfo{author}{\bibfnamefont{T.}~\bibnamefont{Mahood}},
  \bibinfo{journal}{Found. Physics} \textbf{\bibinfo{volume}{29}},
  \bibinfo{pages}{899} (\bibinfo{year}{1999}),
  \eprint{http://chaos.fullerton.edu/~jimw/general/inertia/index.htm}.

\bibitem[{\citenamefont{Ciufolini and Wheeler}(1995)}]{CIU95}
\bibinfo{author}{\bibfnamefont{I.}~\bibnamefont{Ciufolini}} \bibnamefont{and}
  \bibinfo{author}{\bibfnamefont{J.}~\bibnamefont{Wheeler}},
  \emph{\bibinfo{title}{Gravitation and inertia}}
  (\bibinfo{publisher}{Princeton University Press, Princeton, New Jersey},
  \bibinfo{year}{1995}).

\bibitem[{\citenamefont{Dobyns et~al.}(2000)\citenamefont{Dobyns, Rueda, and
  Haisch}}]{DOB00}
\bibinfo{author}{\bibfnamefont{Y.}~\bibnamefont{Dobyns}},
  \bibinfo{author}{\bibfnamefont{A.}~\bibnamefont{Rueda}}, \bibnamefont{and}
  \bibinfo{author}{\bibfnamefont{B.}~\bibnamefont{Haisch}},
  \bibinfo{journal}{Found. Physics} \textbf{\bibinfo{volume}{30}},
  \bibinfo{pages}{59} (\bibinfo{year}{2000}), \eprint{gr-qc/0002069}.

\bibitem[{\citenamefont{Rueda et~al.}(2001)\citenamefont{Rueda, Haisch, and
  Tung}}]{RUE01}
\bibinfo{author}{\bibfnamefont{A.}~\bibnamefont{Rueda}},
  \bibinfo{author}{\bibfnamefont{B.}~\bibnamefont{Haisch}}, \bibnamefont{and}
  \bibinfo{author}{\bibfnamefont{R.}~\bibnamefont{Tung}}
  (\bibinfo{year}{2001}), \eprint{gr-qc/0108026}.

\bibitem[{\citenamefont{Haisch and Rueda}(2001)}]{HAI01}
\bibinfo{author}{\bibfnamefont{B.}~\bibnamefont{Haisch}} \bibnamefont{and}
  \bibinfo{author}{\bibfnamefont{A.}~\bibnamefont{Rueda}}
  (\bibinfo{year}{2001}), \eprint{gr-qc/0106075}.

\bibitem[{\citenamefont{Ibison and Haisch}(1996)}]{IBI96}
\bibinfo{author}{\bibfnamefont{M.}~\bibnamefont{Ibison}} \bibnamefont{and}
  \bibinfo{author}{\bibfnamefont{B.}~\bibnamefont{Haisch}},
  \bibinfo{journal}{Phys. Rev. A} \textbf{\bibinfo{volume}{54}},
  \bibinfo{pages}{2737} (\bibinfo{year}{1996}).

\bibitem[{\citenamefont{Boyer}(1975)}]{BOY75}
\bibinfo{author}{\bibfnamefont{T.~H.} \bibnamefont{Boyer}},
  \bibinfo{journal}{Phys. Rev. D} \textbf{\bibinfo{volume}{11}},
  \bibinfo{pages}{790} (\bibinfo{year}{1975}).

\bibitem[{\citenamefont{de~la Pe\~na and Cetto}(1996)}]{PEN96}
\bibinfo{author}{\bibfnamefont{L.}~\bibnamefont{de~la Pe\~na}}
  \bibnamefont{and} \bibinfo{author}{\bibfnamefont{A.~M.} \bibnamefont{Cetto}},
  \emph{\bibinfo{title}{The Quantum Dice: An Introduction to Stochastic
  Electrodynamics}} (\bibinfo{publisher}{Kluwer Acad. Publ., Dordrecht, the
  Netherlands}, \bibinfo{year}{1996}).

\bibitem[{\citenamefont{Rugh and Zinkernagel}(2002)}]{RUG00}
\bibinfo{author}{\bibfnamefont{S.~E.} \bibnamefont{Rugh}} \bibnamefont{and}
  \bibinfo{author}{\bibfnamefont{H.}~\bibnamefont{Zinkernagel}},
  \bibinfo{journal}{Studies in History and Philosophy of Modern Physics}
  \textbf{\bibinfo{volume}{33}}, \bibinfo{pages}{663} (\bibinfo{year}{2002}),
  \eprint{hep-th/0012253}.

\bibitem[{\citenamefont{Adler et~al.}(1995)\citenamefont{Adler, Casey, and
  Jacob}}]{ADL95}
\bibinfo{author}{\bibfnamefont{R.~J.} \bibnamefont{Adler}},
  \bibinfo{author}{\bibfnamefont{B.}~\bibnamefont{Casey}}, \bibnamefont{and}
  \bibinfo{author}{\bibfnamefont{O.~C.} \bibnamefont{Jacob}},
  \bibinfo{journal}{Am. J. Phys.} \textbf{\bibinfo{volume}{63}},
  \bibinfo{pages}{720} (\bibinfo{year}{1995}).

\bibitem[{\citenamefont{Peebles and Ratra}(2002)}]{PEB02}
\bibinfo{author}{\bibfnamefont{P.}~\bibnamefont{Peebles}} \bibnamefont{and}
  \bibinfo{author}{\bibfnamefont{B.}~\bibnamefont{Ratra}}
  (\bibinfo{year}{2002}), \eprint{astro-ph/0207347}.

\bibitem[{\citenamefont{Webb et~al.}(2001)\citenamefont{Webb, Murphy, Flambaum,
  Dzuba, Barrow, Churchill, Prochaska, and Wolfe}}]{WEB01}
\bibinfo{author}{\bibfnamefont{J.}~\bibnamefont{Webb}},
  \bibinfo{author}{\bibfnamefont{M.}~\bibnamefont{Murphy}},
  \bibinfo{author}{\bibfnamefont{V.}~\bibnamefont{Flambaum}},
  \bibinfo{author}{\bibfnamefont{V.}~\bibnamefont{Dzuba}},
  \bibinfo{author}{\bibfnamefont{J.}~\bibnamefont{Barrow}},
  \bibinfo{author}{\bibfnamefont{C.}~\bibnamefont{Churchill}},
  \bibinfo{author}{\bibfnamefont{J.}~\bibnamefont{Prochaska}},
  \bibnamefont{and} \bibinfo{author}{\bibfnamefont{A.}~\bibnamefont{Wolfe}},
  \bibinfo{journal}{Phys.Rev.Lett.} \textbf{\bibinfo{volume}{87}},
  \bibinfo{pages}{091301} (\bibinfo{year}{2001}), \eprint{astro-ph/0012539}.

\bibitem[{\citenamefont{Fulton and Rohrlich}(1960)}]{FUL60}
\bibinfo{author}{\bibnamefont{Fulton}} \bibnamefont{and}
  \bibinfo{author}{\bibnamefont{Rohrlich}}, \bibinfo{journal}{Annals of
  Physics} \textbf{\bibinfo{volume}{9}}, \bibinfo{pages}{499}
  (\bibinfo{year}{1960}).

\bibitem[{\citenamefont{Bell}(summer 1976)}]{BEL76}
\bibinfo{author}{\bibfnamefont{J.~S.} \bibnamefont{Bell}},
  \bibinfo{journal}{Progress in Scientific Culture}
  \textbf{\bibinfo{volume}{Vol. 1, number 2}} (\bibinfo{year}{summer 1976}).

\bibitem[{\citenamefont{Bell}(1987)}]{BEL87}
\bibinfo{author}{\bibfnamefont{J.~S.} \bibnamefont{Bell}},
  \emph{\bibinfo{title}{Speakable and unspeakable in quantum mechanics}}
  (\bibinfo{publisher}{Cambridge University Press}, \bibinfo{year}{1987}).

\bibitem[{\citenamefont{Liberati et~al.}(2002)\citenamefont{Liberati, Sonego,
  and Visser}}]{LIB02}
\bibinfo{author}{\bibfnamefont{S.}~\bibnamefont{Liberati}},
  \bibinfo{author}{\bibfnamefont{S.}~\bibnamefont{Sonego}}, \bibnamefont{and}
  \bibinfo{author}{\bibfnamefont{M.}~\bibnamefont{Visser}},
  \bibinfo{journal}{Annals Phys.} \textbf{\bibinfo{volume}{298}},
  \bibinfo{pages}{167} (\bibinfo{year}{2002}), \eprint{gr-qc/0107091}.

\bibitem[{\citenamefont{Feynman et~al.}(1964)\citenamefont{Feynman, Leighton,
  and Sands}}]{FEY64}
\bibinfo{author}{\bibfnamefont{R.}~\bibnamefont{Feynman}},
  \bibinfo{author}{\bibfnamefont{R.}~\bibnamefont{Leighton}}, \bibnamefont{and}
  \bibinfo{author}{\bibfnamefont{M.}~\bibnamefont{Sands}},
  \emph{\bibinfo{title}{The Feynman lectures on physics}}
  (\bibinfo{publisher}{Addison-Wesley, Reading, Massachusetts},
  \bibinfo{year}{1964}).

\bibitem[{\citenamefont{Scharnhorst}(1998)}]{SCH98}
\bibinfo{author}{\bibfnamefont{K.}~\bibnamefont{Scharnhorst}},
  \bibinfo{journal}{Annalen Phys.} \textbf{\bibinfo{volume}{7}},
  \bibinfo{pages}{700} (\bibinfo{year}{1998}), \eprint{hep-th/9810221}.

\bibitem[{\citenamefont{Lobo and Crawford}()}]{LOB02}
\bibinfo{author}{\bibfnamefont{F.}~\bibnamefont{Lobo}} \bibnamefont{and}
  \bibinfo{author}{\bibfnamefont{P.}~\bibnamefont{Crawford}},
  \eprint{gr-qc/0204038}.

\bibitem[{\citenamefont{Magueijo et~al.}(2002)\citenamefont{Magueijo, Barrow,
  and Sandvik}}]{MAG02}
\bibinfo{author}{\bibfnamefont{J.}~\bibnamefont{Magueijo}},
  \bibinfo{author}{\bibfnamefont{J.~D.} \bibnamefont{Barrow}},
  \bibnamefont{and} \bibinfo{author}{\bibfnamefont{H.~B.}
  \bibnamefont{Sandvik}}, \bibinfo{journal}{Phys.Lett. B}
  \textbf{\bibinfo{volume}{549}}, \bibinfo{pages}{284} (\bibinfo{year}{2002}),
  \eprint{astro-ph/0202374}.

\bibitem[{\citenamefont{Roll et~al.}(1964)\citenamefont{Roll, Krotkov, and
  Dicke}}]{ROL64}
\bibinfo{author}{\bibfnamefont{P.}~\bibnamefont{Roll}},
  \bibinfo{author}{\bibfnamefont{R.}~\bibnamefont{Krotkov}}, \bibnamefont{and}
  \bibinfo{author}{\bibfnamefont{R.}~\bibnamefont{Dicke}},
  \bibinfo{journal}{Ann. Phys. (N.Y.)} \textbf{\bibinfo{volume}{26}},
  \bibinfo{pages}{442} (\bibinfo{year}{1964}).

\bibitem[{\citenamefont{Braginsky and Panov}(1972)}]{BRA72}
\bibinfo{author}{\bibfnamefont{V.}~\bibnamefont{Braginsky}} \bibnamefont{and}
  \bibinfo{author}{\bibfnamefont{V.}~\bibnamefont{Panov}},
  \bibinfo{journal}{Sov. Phys. JEPT} \textbf{\bibinfo{volume}{43}},
  \bibinfo{pages}{463} (\bibinfo{year}{1972}).

\bibitem[{\citenamefont{Adelberg}(1994)}]{ADE94}
\bibinfo{author}{\bibfnamefont{E.~G.} \bibnamefont{Adelberg}},
  \bibinfo{journal}{Class. Quant. Grav.} \textbf{\bibinfo{volume}{11}},
  \bibinfo{pages}{A9} (\bibinfo{year}{1994}).

\bibitem[{\citenamefont{Nordtvedt}(2002)}]{NOR02}
\bibinfo{author}{\bibfnamefont{K.}~\bibnamefont{Nordtvedt}}
  (\bibinfo{year}{2002}), \eprint{gr-qc/0212044}.

\bibitem[{\citenamefont{Gordon}(1923)}]{GOR23}
\bibinfo{author}{\bibfnamefont{W.}~\bibnamefont{Gordon}},
  \bibinfo{journal}{Ann. Phys. Leipzig} \textbf{\bibinfo{volume}{72}},
  \bibinfo{pages}{421} (\bibinfo{year}{1923}).

\bibitem[{\citenamefont{Visser et~al.}(2002)\citenamefont{Visser, Barcel\'o,
  and Liberati}}]{VIS01}
\bibinfo{author}{\bibfnamefont{M.}~\bibnamefont{Visser}},
  \bibinfo{author}{\bibfnamefont{C.}~\bibnamefont{Barcel\'o}},
  \bibnamefont{and} \bibinfo{author}{\bibfnamefont{S.}~\bibnamefont{Liberati}},
  \bibinfo{journal}{Gen.Rel.Grav.} \textbf{\bibinfo{volume}{34}},
  \bibinfo{pages}{1719} (\bibinfo{year}{2002}), \eprint{gr-qc/0111111}.

\bibitem[{\citenamefont{Bondi and Samuel}(1996)}]{BON96}
\bibinfo{author}{\bibfnamefont{H.}~\bibnamefont{Bondi}} \bibnamefont{and}
  \bibinfo{author}{\bibfnamefont{J.}~\bibnamefont{Samuel}}
  (\bibinfo{year}{1996}), \eprint{gr-qc/9607009}.

\bibitem[{\citenamefont{Bohm and Aharonov}(1957)}]{BHO57}
\bibinfo{author}{\bibfnamefont{D.}~\bibnamefont{Bohm}} \bibnamefont{and}
  \bibinfo{author}{\bibfnamefont{Y.}~\bibnamefont{Aharonov}},
  \bibinfo{journal}{Phys. Rev.} \textbf{\bibinfo{volume}{108}},
  \bibinfo{pages}{1070} (\bibinfo{year}{1957}).

\bibitem[{\citenamefont{Joos and Zeh}(1985)}]{JOO85}
\bibinfo{author}{\bibfnamefont{E.}~\bibnamefont{Joos}} \bibnamefont{and}
  \bibinfo{author}{\bibfnamefont{H.}~\bibnamefont{Zeh}}, \bibinfo{journal}{Z.
  Phys. B} \textbf{\bibinfo{volume}{59}}, \bibinfo{pages}{223}
  (\bibinfo{year}{1985}).

\bibitem[{\citenamefont{Bell}(1964)}]{BEL64}
\bibinfo{author}{\bibfnamefont{J.}~\bibnamefont{Bell}},
  \bibinfo{journal}{Physics} \textbf{\bibinfo{volume}{1}}, \bibinfo{pages}{195}
  (\bibinfo{year}{1964}).

\bibitem[{\citenamefont{Clausser et~al.}(1969)\citenamefont{Clausser, Horne,
  Shimony, and Holt}}]{CLA69}
\bibinfo{author}{\bibfnamefont{J.~F.} \bibnamefont{Clausser}},
  \bibinfo{author}{\bibfnamefont{M.}~\bibnamefont{Horne}},
  \bibinfo{author}{\bibfnamefont{A.}~\bibnamefont{Shimony}}, \bibnamefont{and}
  \bibinfo{author}{\bibfnamefont{R.}~\bibnamefont{Holt}},
  \bibinfo{journal}{Phys. Rev. Lett.} \textbf{\bibinfo{volume}{26}},
  \bibinfo{pages}{880} (\bibinfo{year}{1969}).

\bibitem[{\citenamefont{Wignwe}(1970)}]{WIG70}
\bibinfo{author}{\bibfnamefont{E.}~\bibnamefont{Wignwe}}, \bibinfo{journal}{Am.
  J. Phys.} \textbf{\bibinfo{volume}{38}}, \bibinfo{pages}{1005}
  (\bibinfo{year}{1970}).

\bibitem[{\citenamefont{Aspect et~al.}(1982)\citenamefont{Aspect, Dalibard, and
  Roger}}]{ASP82}
\bibinfo{author}{\bibnamefont{Aspect}},
  \bibinfo{author}{\bibnamefont{Dalibard}}, \bibnamefont{and}
  \bibinfo{author}{\bibnamefont{Roger}}, \bibinfo{journal}{Phys. Rev. Lett.}
  \textbf{\bibinfo{volume}{49}}, \bibinfo{pages}{1804} (\bibinfo{year}{1982}).

\bibitem[{\citenamefont{Bohm}(1952)}]{BHO52}
\bibinfo{author}{\bibfnamefont{D.}~\bibnamefont{Bohm}}, \bibinfo{journal}{Phys.
  Rev.} \textbf{\bibinfo{volume}{85}}, \bibinfo{pages}{166 and 180}
  (\bibinfo{year}{1952}).

\bibitem[{\citenamefont{Pearle}(1999)}]{PEA99}
\bibinfo{author}{\bibfnamefont{P.}~\bibnamefont{Pearle}},
  \emph{\bibinfo{title}{Open Systems and Measurement in Relativistic Quantum
  Theory}} (\bibinfo{publisher}{Springer-Verlag}, \bibinfo{year}{1999}), chap.
  \bibinfo{chapter}{Collapse models}, \eprint{quant-ph/9901077}.

\bibitem[{\citenamefont{Pearle}(1979)}]{PEA79}
\bibinfo{author}{\bibfnamefont{P.}~\bibnamefont{Pearle}},
  \bibinfo{journal}{Int. Jour. Theor. Phys.} \textbf{\bibinfo{volume}{48}},
  \bibinfo{pages}{489} (\bibinfo{year}{1979}).

\bibitem[{\citenamefont{Scharf}(1995)}]{SCH95}
\bibinfo{author}{\bibfnamefont{G.}~\bibnamefont{Scharf}},
  \emph{\bibinfo{title}{Finite Quantum Electrodynamics}}
  (\bibinfo{publisher}{Springer-Verlag}, \bibinfo{year}{1995}).

\bibitem[{\citenamefont{Schroedinger}(1926)}]{SCH}
\bibinfo{author}{\bibfnamefont{E.}~\bibnamefont{Schroedinger}},
  \bibinfo{journal}{Annalen der Physik} \textbf{\bibinfo{volume}{79}},
  \bibinfo{pages}{489} (\bibinfo{year}{1926}).

\bibitem[{\citenamefont{Unruh}(1995)}]{UNR95}
\bibinfo{author}{\bibfnamefont{W.~G.} \bibnamefont{Unruh}},
  \bibinfo{journal}{Phys. Rev. D} \textbf{\bibinfo{volume}{51}},
  \bibinfo{pages}{2827} (\bibinfo{year}{1995}), \eprint{gr-qc/9409008}.

\bibitem[{\citenamefont{Barcel\'o and Visser}(2002)}]{BAR02c}
\bibinfo{author}{\bibfnamefont{C.}~\bibnamefont{Barcel\'o}} \bibnamefont{and}
  \bibinfo{author}{\bibfnamefont{M.}~\bibnamefont{Visser}},
  \bibinfo{journal}{Int.J.Mod.Phys. D} \textbf{\bibinfo{volume}{11}},
  \bibinfo{pages}{1553} (\bibinfo{year}{2002}), \eprint{gr-qc/0205066}.

\bibitem[{\citenamefont{Rañada}(2002)}]{RAN02}
\bibinfo{author}{\bibfnamefont{A.~F.} \bibnamefont{Rañada}}
  (\bibinfo{year}{2002}), \eprint{gr-qc/0211052}.

\bibitem[{\citenamefont{Ellis}(2002)}]{ELL01}
\bibinfo{author}{\bibfnamefont{G.~F.~R.} \bibnamefont{Ellis}},
  \bibinfo{journal}{Int.J.Mod.Phys. A} \textbf{\bibinfo{volume}{17}},
  \bibinfo{pages}{2667} (\bibinfo{year}{2002}), \eprint{gr-qc/0102017}.

\bibitem[{\citenamefont{Ryder}(1985)}]{RYD85}
\bibinfo{author}{\bibfnamefont{L.}~\bibnamefont{Ryder}},
  \emph{\bibinfo{title}{Quantum field theory}} (\bibinfo{publisher}{Cambridge
  University Press}, \bibinfo{year}{1985}).

\end{thebibliography}

\newpage

\end{document}